\documentclass[a4paper,11pt]{article}
\pdfoutput=1 

\usepackage{jheppub} 

\usepackage[T1]{fontenc} 
\usepackage[dvipsnames]{xcolor}
\usepackage{relsize}
\usepackage{tikz}

\def\res{\mathop{\rm Res}}

\title{\boldmath From bulk loops to boundary large-N expansion}

\author{Dmitry Ponomarev}

\affiliation{Texas A\&M University, College Station, TX 77843, USA}
\affiliation{Lebedev Physical Institute, Moscow, 119991, Russia}

\emailAdd{ponomarev@lpi.ru}


\abstract{We study the analytic structure of loop Witten diagrams in Euclidean AdS represented by their conformal partial wave expansions. We show that, as in flat space, amplitude's singularities are associated with non-trivial cuts of the diagram and factorize into products of the coefficient functions for the subdiagrams resulting from these cuts. We consider an example of a one-loop four-point diagram in detail and then briefly discuss how the procedure can be extended to more general diagrams. Finally, we  show that this analysis reproduces simple relations that follow from the large-$N$ considerations on the boundary.

}

\begin{document} 

\begin{flushright}\small{MI-TH-1935}\end{flushright}
\maketitle
\flushbottom

\tikzset{
pic1/.pic={
\draw[thick] (0,0) circle [radius=3];
\filldraw [fill=Peach,draw=black, thick]  (-1.25,0) circle [radius=0.75];
\filldraw [fill=Peach,draw=black, thick] (1.25,0) circle [radius=0.75];
\draw[thick, RoyalBlue]  (-1.25,0) ++(135:0.75) --(150:3);
\draw[thick, RoyalBlue]  (-1.25,0) ++(225:0.75) --(210:3);
\draw[thick, RoyalBlue]  (1.25,0) ++(45:0.75) --(30:3);
\draw[thick, RoyalBlue]  (1.25,0) ++(-45:0.75) --(-30:3);
\draw[thick, RoyalBlue] (-1.25,0) ++(45:0.75) .. controls ({-1.25+1.3*cos(45)},{0+1.3*sin(45)})  and ({1.25+1.3*cos(135)},{0+1.3*sin(135)}) .. ({1.25+0.75*cos(135)},{0+0.75*sin(135)});
\draw[thick, RoyalBlue] ({-1.25+0.75*cos(-45)},{0+0.75*sin(-45)}) .. controls ({-1.25+1.3*cos(-45)},{0+1.3*sin(-45)}) and ({1.25+1.3*cos(-135)},{0+1.3*sin(-135)}) .. ({1.25+0.75*cos(-135)},{0+0.75*sin(-135)});
\draw (-1.25,0) node {L};
\draw (1.25,0) node {R};
}
}

\tikzset{
pic2/.pic={
\draw[thick] (0,0) circle [radius=3];
\filldraw [fill=Peach,draw=black, thick]  (-1.25,0) circle [radius=0.75];
\filldraw [fill=Peach,draw=black, thick] (1.25,0) circle [radius=0.75];
\draw[thick, RoyalBlue]  (-1.25,0) ++(135:0.75) --(150:3);
\draw[thick, RoyalBlue]  (-1.25,0) ++(225:0.75) --(210:3);
\draw[thick, RoyalBlue]  (1.25,0) ++(45:0.75) --(30:3);
\draw[thick, RoyalBlue]  (1.25,0) ++(-45:0.75) --(-30:3);
\draw[thick, RoyalBlue, rounded corners=1pt]  ({-1.25+0.75*cos(45)},{0+0.75*sin(45)}) -- (0,3) -- ({1.25+0.75*cos(135)},{0+0.75*sin(135)});
\draw[thick, RoyalBlue, rounded corners=1pt]  ({-1.25+0.75*cos(-45)},{0+0.75*sin(-45)}) -- (0,-3) -- ({1.25+0.75*cos(-135)},{0+0.75*sin(-135)});
\draw (-1.25,0) node {L};
\draw (1.25,0) node {R};
\draw (0,3) node[anchor=south] {$\nu_1$,$P_1$};
\draw (0,-3) node[anchor=north] {$\nu_2$,$P_2$};
\draw (-3,0) node[anchor=east] {$
\mathlarger{\int}
\left\{\begin{array}{c}
d\nu_1\\
 d\nu_2\\
dP_1\\
dP_2
\end{array}\right\}$
};
}
}

\tikzset{
pic3/.pic={
\draw[thick] (0,0) circle [radius=3];
\draw[thick, RoyalBlue, rounded corners=1pt] ({3*cos(150)},{3*sin(150)}) --  ({3*cos(110)},{3*sin(110)}) --  ({3*cos(-150)},{3*sin(-150)}) -- cycle;
\draw[thick, OliveGreen, rounded corners=1pt] ({3*cos(110)},{3*sin(110)}) --  ({3*cos(90)},{3*sin(90)}) --  ({3*cos(-90)},{3*sin(-90)}) -- cycle;
\draw[thick, OliveGreen, rounded corners=1pt] ({3*cos(90)},{3*sin(90)}) --  ({3*cos(-90)},{3*sin(-90)}) --  ({3*cos(-50)},{3*sin(-50)}) -- cycle;
\draw[thick, RoyalBlue, rounded corners=1pt] ({3*cos(30)},{3*sin(30)}) --  ({3*cos(-30)},{3*sin(-30)}) --  ({3*cos(-50)},{3*sin(-50)}) -- cycle;
\draw (0,3) node[anchor=south] {$\nu_1$,$P_1$};
\draw (0,-3) node[anchor=north] {$\nu_2$,$P_2$};
\draw ({3*cos(110)},{3*sin(110)}) node[anchor=south east] {$\nu_{\rm L}$,$P_{\rm L}$};
\draw ({3*cos(-50)},{3*sin(-50)}) node[anchor=north west] {$\nu_{\rm R}$,$P_{\rm R}$};\draw (-3,0) node[anchor=east] {$
\mathlarger{\int}
\left\{\begin{array}{c}
d\nu_1\\
 d\nu_2\\
 d\nu_{\rm L}\\
 d\nu_{\rm R}\\
dP_1\\
dP_2\\
dP_{\rm L}\\
dP_{\rm R}
\end{array}\right\}$
};
}
}

\tikzset{
pic4/.pic={
\draw[thick] (0,0) circle [radius=3];
\draw[thick, RoyalBlue, rounded corners=1pt] ({3*cos(150)},{3*sin(150)}) --  ({3*cos(110)},{3*sin(110)}) --  ({3*cos(-150)},{3*sin(-150)}) -- cycle;
\draw[thick, RoyalBlue, rounded corners=1pt] ({3*cos(30)},{3*sin(30)}) --  ({3*cos(-30)},{3*sin(-30)}) --  ({3*cos(110)},{3*sin(110)}) -- cycle;
\draw ({3*cos(110)},{3*sin(110)}) node[anchor=south east] {$\nu_{\rm L}$,$P_{\rm L}$};
\draw (-3,0) node[anchor=east] {$
\mathlarger{\int}
\left\{\begin{array}{c}
d\nu_1\\
 d\nu_2\\
 d\nu_{\rm L}\\
dP_{\rm L}
\end{array}\right\}$
};
}
}

\tikzset{
pic5/.pic={
\draw[thick] (0,0) circle [radius=3];
\filldraw [fill=Peach,draw=black, thick]  (0,0) circle [radius=0.75];
\draw[thick, RoyalBlue] ({0.75*cos(45)},{0.75*sin(45)}) -- ({3*cos(45)},{3*sin(45)});
\draw[thick, RoyalBlue] ({0.75*cos(135)},{0.75*sin(135)}) -- ({3*cos(135)},{3*sin(135)});
\draw[thick, RoyalBlue] ({0.75*cos(-45)},{0.75*sin(-45)}) -- ({3*cos(-45)},{3*sin(-45)});
\draw[thick, RoyalBlue] ({0.75*cos(-135)},{0.75*sin(-135)}) -- ({3*cos(-135)},{3*sin(-135)});
\draw (0,0) node {L};
}
}

\tikzset{
pic6/.pic={
\draw[thick] (0,0) circle [radius=3];
\filldraw [fill=Peach,draw=black, thick]  (0,0) circle [radius=0.75];
\draw[thick, RoyalBlue] ({0.75*cos(45)},{0.75*sin(45)}) -- ({3*cos(45)},{3*sin(45)});
\draw[thick, RoyalBlue] ({0.75*cos(135)},{0.75*sin(135)}) -- ({3*cos(135)},{3*sin(135)});
\draw[thick, RoyalBlue] ({0.75*cos(-45)},{0.75*sin(-45)}) -- ({3*cos(-45)},{3*sin(-45)});
\draw[thick, RoyalBlue] ({0.75*cos(-135)},{0.75*sin(-135)}) -- ({3*cos(-135)},{3*sin(-135)});
\draw (0,0) node {R};
}
}

\section{Introduction}
\label{sec:intro}

The AdS/CFT correspondence is a remarkable duality that relates quantum field theories in anti de Sitter space and conformal field theories on its boundary \cite{Maldacena:1997re,Witten:1998qj,Gubser:1998bc}. Mathematically, it is stated as the equality between the bulk path integral with the properly set boundary conditions and the generating function of the CFT correlators on the boundary. The AdS/CFT correspondence has attracted significant interest in recent years as it provides new tools for addressing important and challenging issues of quantum gravity and strongly coupled systems. Probably, the most studied regime of the AdS/CFT correspondence is the one in which the bulk theory is weakly coupled, while the boundary theory has many degrees of freedom. In this regime the bulk loop expansion translates into the $1/N$ expansion of the large-$N$ CFT. 

On the CFT side, the large-$N$ expansion can be studied using various methods. In particular, in certain cases the associated diagrammatic expansions are available. Alternatively, one can use the  large-$N$ bootstrap, which amounts to solving the crossing equations perturbatively in  $1/N$. In this approach one starts with the CFT data of  mean field theory, which is of order $O(N^0)$ and solves the crossing equations identically. Next, one gives $O(1/N^2)$ corrections to the CFT data\footnote{Here we give the orders in $1/N$ as they appear in models with operators in the adjoint representation of some internal algebra with large $N$. For vector models $1/N^2$ should be replaced with $1/N$.}. By imposing crossing to  order $O(1/N^2)$, one finds constraints on these corrections. These define four-point correlators at order $O(1/N^2)$, which, using holography, can be reinterpreted as bulk tree-level diagrams. For comprehensive analysis at this order in the holographic context see \cite{Heemskerk:2009pn,Alday:2017gde}. 
Proceeding further, one finds, that the CFT data at order $O(1/N^2)$ sources $O(1/N^4)$ contributions to the crossing equations. To satisfy crossing at this order, the CFT data should receive $O(1/N^4)$ corrections.
 This procedure should be repeated iteratively, thus reproducing the CFT data order by order.  
Impressive progress in this direction was achieved in recent years \cite{Alday:2016njk,Aharony:2016dwx,Caron-Huot:2017vep,Alday:2017gde,Alday:2017xua,Aprile:2017bgs,Alday:2017vkk,Aprile:2017qoy,Aharony:2018npf,Alday:2018pdi,Ghosh:2018bgd,Alday:2018kkw,Ponomarev:2019ltz,Alday:2019qrf,Alday:2019clp}. In particular, the  CFT data to  order $O(1/N^4)$ in different theories was computed. 
Though, these results were derived from large-$N$ considerations in the CFT, assuming holography, these are also regarded as one-loop computations in the bulk.

In turn, on the bulk side, despite some direct computations of loop amplitudes are available, the literature on the subject remains limited. First progress was made in \cite{Penedones:2010ue,Fitzpatrick:2011hu,Fitzpatrick:2011dm}, where bubble diagrams were computed in the Mellin representation. Later, further results were obtained in different representations \cite{Cardona:2017tsw,Giombi:2017hpr,Yuan:2017vgp,Yuan:2018qva,Bertan:2018khc,Liu:2018jhs,Bertan:2018afl}.  In the course of this work it was found that amplitudes in AdS have a specific analytic structure, similar to the analytic structure of  amplitudes in flat space. In particular, locations of poles in the Mellin amplitude for the bubble diagram were identified in \cite{Penedones:2010ue}. Then, this amplitude was computed exactly in \cite{Fitzpatrick:2011hu,Fitzpatrick:2011dm} and contributions to the conformal block decomposition associated with the singular part of the Mellin amplitude were found. It was further shown, that the conformal block coefficients in this decomposition factorize into the OPE coefficients associated with tree-level diagrams, obtained by cutting two lines in  the original bubble diagram --- which is exactly the relation that one expects from large-$N$ considerations on the boundary.  Finally, by taking the flat space limit, this factorization property was related to unitarity. The analytic structure of more general amplitudes was later studied in Mellin space  \cite{Yuan:2017vgp,Yuan:2018qva} and similar factorization patterns were observed.

In the present paper we will further investigate the analytic structure of loop amplitudes in AdS and show how factorization of amplitude's singularities translates into familiar relations from the large-$N$ bootstrap. We perform our analysis using the conformal partial wave expansion for bulk amplitudes, because this representation makes the connection with the CFT data on the boundary straightforward. Besides that, the conformal partial wave expansion seems to be more suitable for treating higher-spin theories in the bulk, for which Mellin amplitudes degenerate \cite{Taronna:2016ats,Bekaert:2016ezc,Ponomarev:2017qab,Rastelli:2017udc}.

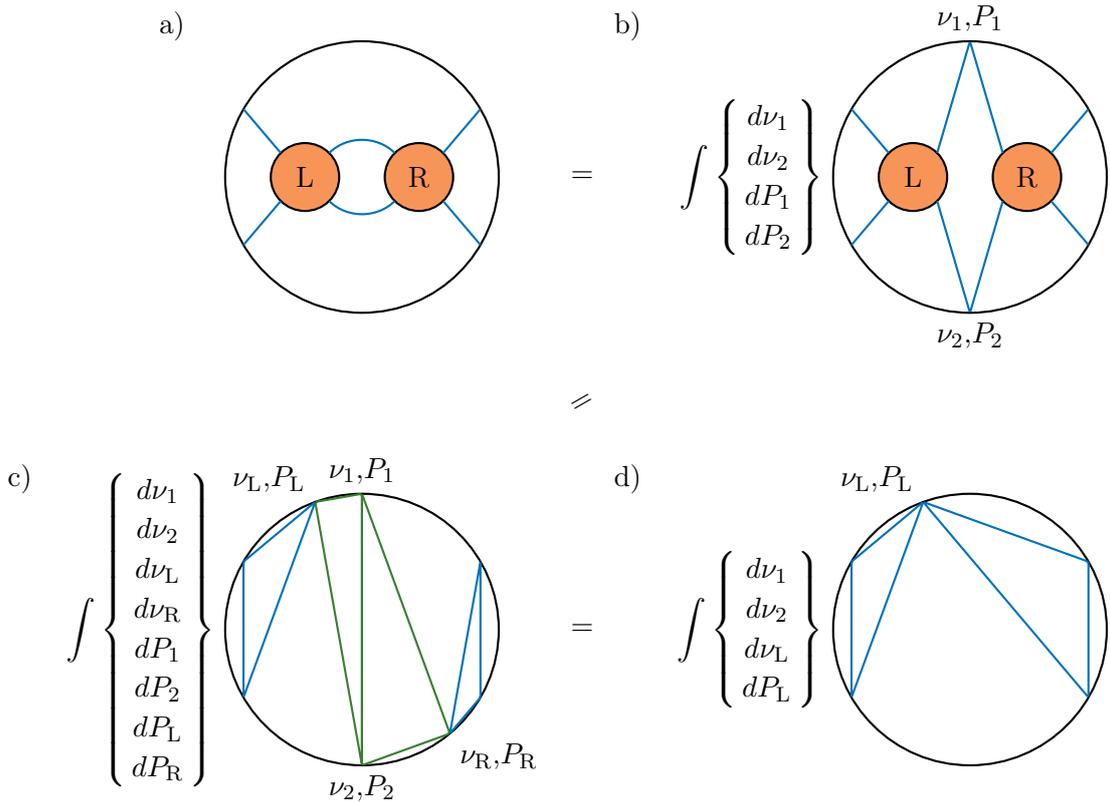
\begin{figure}
\label{fig:1}
\centering
\begin{tikzpicture}
\draw (0,0)  pic[scale=0.6] {pic1};
\draw (8,0) pic[scale=0.6] {pic2};
\draw (0,-6) pic[scale=0.6] {pic3};
\draw (8,-6) pic[scale=0.6] {pic4};
\draw (2.9,0) node {$=$};
\draw (2.9,-6) node {$=$};
\draw (2.9,-3) node  [rotate=30] {$=$};
\draw (-2.5,2) node {a)};
\draw ({-4.5+8},{2+0}) node {b)};
\draw ({-4.5+0},{2-6}) node {c)};
\draw ({-4.5+8},{2-6}) node {d)};
\end{tikzpicture}
\caption{This figure illustrates the key steps of section \ref{sec:loopfromtree}. First, we use the split representation for a pair of propagators to represent loop diagram a) in the from b). Blobs L and R refer to generic bulk processes, which, for simplicity, we consider to be tree diagrams.
 Next, we substitute the conformal partial wave expansions for the subdiagrams  into b), which gives c). In c) triangles denote the properly normalized conformally-invariant three-point structures. Finally, we evaluate the bubble integral for the structures highlighted in green, which leads to the conformal partial wave expansion for the original loop diagram d).}
\end{figure}

\begin{figure}
\centering
\begin{tikzpicture}
\draw (0,0) pic[scale=0.6] {pic1};
\draw (4.8,0) pic[scale=0.6] {pic5};
\draw (9.5,0) pic[scale=0.6] {pic6};
\draw (-2.5,0) node {d.t.s $\Bigg[$};
\draw(2.3,0) node {$\Bigg]\;\sim$};
\draw (7.2,0) node {$\times$};
\end{tikzpicture}
\caption{Using representation d) from Fig \ref{fig:1}
 for the loop diagram, we study the analytic structure of the coefficient function of its conformal partial wave expansion in section \ref{sec:sing}. We find that this coefficient function has singularities at double-trace locations, which, moreover, factorize into coefficient functions for subdiagrams. On the figure ''d.t.s.'' refers to double-trace singularities.}
 \label{newlabel}
\end{figure}
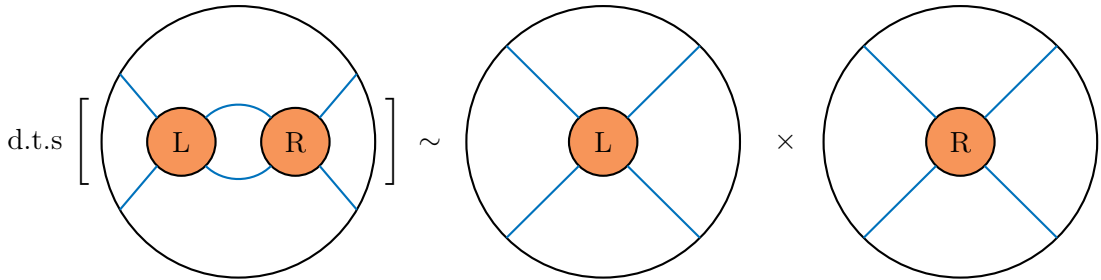

The key steps of our computation are as follows, see Fig \ref{fig:1}. For a given cut of the loop amplitude, we factorize each propagator that we are going to cut using the split representation. This brings the original amplitude into a form of an integrated product of off-shell amplitudes for the subdiagrams resulting from the cut. Next, we use the conformal partial wave expansion for the subdiagrams in a suitable channel and integrate over auxiliary boundary points introduced by the split representation. To this end we use the bubble integral formula \cite{Dobrev:1976vr,Karateev:2018oml} iteratively until the space-time dependence reduces to a single partial wave. This eventually yields the conformal partial wave expansion for the original loop diagram with the coefficient function given as an  integral over spectral parameters of propagators. Analytic structure of these integrals is then studied using the standard methods, Fig. \ref{newlabel},  see \cite{smatrix,Yuan:2018qva} for review.

As a result, we find that the singular part of the original amplitude associated with a given cut can be computed as a product of on-shell amplitudes for subdiagrams, integrated over the on-shell phase space of the particles on the cut lines --- the same way as it happens in flat space. We then give diagrammatic rules to compute these singularities --- the AdS version of the Cutkosky rules. Next, we show that the previously found relation between the amplitudes for the diagram and its subdiagrams is consistent with the standard large-$N$ considerations on the boundary. We carry out the above analysis in detail for the case of a double-particle cut of a four-point amplitude and then show how this approach can be applied in more general situations.

The paper is organized as follows. In section \ref{sec:tree} we review the necessary background material for computing Witten diagrams in the form of the conformal partial wave expansion. Next, in section \ref{sec:loopfromtree} we express a one-loop four-point amplitude with a non-trivial double-cut in a given channel in terms of off-shell amplitudes associated with the subdiagrams produced by the cut. In section \ref{sec:sing} we use the standard methods to analyze the analytic structure of the integrals we encountered in the previous section. Agreement with the large-$N$ analysis is established in section \ref{sec:largeN}. Then, in section \ref{sec:gen} we briefly discuss various consequences and generalizations of the presented approach. Finally, we give our conclusions in section \ref{sec:conc}. In Appendix \ref{App:Ex} we illustrate how the standard analysis of integral's singularities works with a simple example.

\section{Amplitudes in AdS: the basics}
\label{sec:tree}

In this section we review how the conformal block decomposition for tree-level Witten diagrams in Euclidean AdS${}_{d+1}$ can be derived. 
For simplicity here we will focus on scalar fields only. For more details we refer the reader to  \cite{Penedones:2010ue,Paulos:2011ie,Costa:2014kfa}.

\subsection{Two-point functions}
Witten diagrams involve two types of propagators: bulk-to-boundary propagators for external lines and bulk-to-bulk propagators for internal lines. The canonically normalized bulk-to-boundary propagator for a scalar field of dimension $\Delta$ is given by
\begin{equation}
\label{2jun2}
\Pi_\Delta(X,P)=\frac{{\cal C}_{\Delta}}{(-2 P\cdot X)^\Delta}, \qquad {\cal C}_\Delta = \frac{\Gamma(\Delta)}{2\pi^h \Gamma(\Delta+1-h)}, \qquad h\equiv \frac{d}{2}.
\end{equation}
Throughout the paper we use the ambient space formalism for AdS and CFT, $X$ refer to bulk, while $Q$ and $P$ to boundary points.

It will be convenient to deal with the bulk-to-bulk propagators in the so-called split representation \cite{Penedones:2010ue,Paulos:2011ie,Costa:2014kfa}\footnote{See also \cite{Fronsdal:1974ew,Dobrev:1998md,Leonhardt:2003qu,Leonhardt:2003sn} for earlier closely related results.}. To this end one first defines harmonic functions
\begin{equation}
\label{2jun4}
\Omega_\nu (X_1,X_2) \equiv \frac{\nu^2}{\pi} \int dP \Pi_{h+i\nu}(X_1,P)\Pi_{h-i\nu} (X_2,P), \quad
\Omega_\nu(X_1,X_2)=\Omega_{-\nu}(X_1,X_2).
\end{equation}
They satisfy the free equation of motion identically
\begin{equation}
\label{2jun4x1}
\left(\nabla_1^2+h^2+\nu^2\right)\Omega_\nu(X_1,X_2)=0
\end{equation}
and are used as a basis in the space of bulk two-point functions. In particular, one has 
\begin{equation}
\label{2jun5}
\delta(X_1,X_2)=\int_{-\infty}^\infty d\nu \Omega_\nu (X_1,X_2).
\end{equation}
Employing (\ref{2jun4x1}) and (\ref{2jun5}), it is straightforward to solve 
\begin{equation}
\label{2jun1}
(\nabla_1^2 -m^2)\Pi_\Delta(X_1,X_2)=-\delta(X_1,X_2), \qquad m^2=\Delta(\Delta-d)
\end{equation}
for the bulk-to-bulk propagator as
\begin{equation}
\label{2jun6}
\Pi_\Delta (X_1,X_2)=\int d\nu \frac{1}{\nu^2 + (\Delta-h)^2} \Omega_{\nu}(X_1,X_2).
\end{equation}

On the CFT side, it is conventional to use normalization 
\begin{equation}
\label{2jun6x1}
\langle {\cal O}_\Delta(P_1) {\cal O}_\Delta(P_2)\rangle = \frac{1}{P_{12}^\Delta}, \qquad
P_{12}\equiv -2P_1\cdot P_2. 
\end{equation}
It is not hard to see that in the boundary limit of the bulk-to-boundary propagator (\ref{2jun2})  we arrive at (\ref{2jun6x1}) times ${\cal C}_\Delta$. 
To account  for this difference between bulk and boundary normalizations, before comparing with the CFT side, Witten diagrams have to be divided by a factor of ${\cal C}_{\Delta_i}^{1/2}$ for each external line of the diagram. Note that ${\cal C}_{\Delta}>0$ for $\Delta$
above the unitarity bound, so square roots of ${\cal C}_{\Delta}$ in relevant theories are defined unambiguously.

\subsection{Three-point functions}
A three-point Witten diagram for scalars of dimensions $\Delta_i$ is defined by
\begin{equation}
\label{2jun9}
\begin{split}
A_3(P_1,P_2,P_3)=g_3 \int dX \Pi_{\Delta_1}(X,P_1)\Pi_{\Delta_2}(X,P_2)\Pi_{\Delta_3}(X,P_3),
\end{split}
\end{equation}
where $g_3$ is the bulk coupling constant. Evaluating the integral, one finds  \cite{Freedman:1998tz}
\begin{equation}
\label{2jun9x1}
\begin{split}
A_3(P_1,P_2,P_3)= g_3 b(\Delta_1,\Delta_2,\Delta_3) [ {\cal O}_{\Delta_1}(P_1) {\cal O}_{\Delta_2}(P_2) {\cal O}_{\Delta_3}(P_3) ],
\end{split}
\end{equation}
where
\begin{equation}
\label{2jun10}
\begin{split}
&b(\Delta_1,\Delta_2,\Delta_3)\\
&\quad=
{\cal C}_{\Delta_1}{\cal C}_{\Delta_2}{\cal C}_{\Delta_3}
\frac{\pi^h\Gamma\left(\frac{\Delta_1+\Delta_2+\Delta_3-d}{2} \right)
\Gamma\left(\frac{\Delta_1+\Delta_2-\Delta_3}{2} \right)
\Gamma\left(\frac{\Delta_3+\Delta_1-\Delta_2}{2} \right)
\Gamma\left(\frac{\Delta_3+\Delta_2-\Delta_1}{2} \right)
}{2 \Gamma(\Delta_1)\Gamma(\Delta_2)\Gamma(\Delta_3)}
\end{split}
\end{equation}
and 
\begin{equation}
\label{2jun10x1}
[ {\cal O}_{\Delta_1}(P_1) {\cal O}_{\Delta_2}(P_2) {\cal O}_{\Delta_3}(P_3)]
\equiv \frac{1}{P_{12}^{\frac{\Delta_1+\Delta_2-\Delta_3}{2}}
P_{23}^{\frac{\Delta_2+\Delta_3-\Delta_1}{2}}
P_{31}^{\frac{\Delta_3+\Delta_1-\Delta_2}{2}}
}.
\end{equation}
Here we use square brackets to denote conformally invariant three-point functions with a unit normalization. These are not three-point correlators of a CFT as they do not include the OPE coefficients.

\subsection{Partial waves and conformal blocks}

Four-point functions will be presented in the form of the conformal partial wave expansion, which can then be reduced to the conformal block decomposition by evaluating the contour integral. Originally, this approach was developed in  \cite{Mack:1974jjo,Mack:1974sa,Dobrev:1975ru,Dobrev:1977qv}.
 Below we fix conventions and give some relevant formulae, see  \cite{Simmons-Duffin:2017nub,Karateev:2018oml} for further details and references.

Conformal partial waves are defined by 
\begin{equation}
\label{2jun11}
\Psi^{\Delta_i}_\Delta(P_i)\equiv
\int dP_0 
[ {\cal O}_{\Delta_1}(P_1) {\cal O}_{\Delta_2}(P_2) {\cal O}_{\Delta}(P_0) ]
[ {\cal O}_{\tilde\Delta}(P_0) {\cal O}_{\Delta_3}(P_3) {\cal O}_{\Delta_4}(P_4) ].
\end{equation}
Each conformal partial wave is a linear combination of a conformal block and its shadow partner
\begin{equation}
\label{2jun12}
\Psi^{\Delta_i}_{\Delta}(P_i) = S_{\tilde\Delta}^{\Delta_3,\Delta_4} G^{\Delta_i}_\Delta(P_i)+
S_{\Delta}^{\Delta_1,\Delta_2} G^{\Delta_i}_{\tilde\Delta}(P_i), \qquad \tilde \Delta \equiv d-\Delta,
\end{equation}
where
\begin{equation}
\label{2jun13}
S_{\Delta}^{\Delta_1,\Delta_2}=\frac{\pi^h \Gamma(\Delta - h) \Gamma\left(\frac{\tilde\Delta+\Delta_1-\Delta_2}{2} \right)
\Gamma\left(\frac{\tilde\Delta+\Delta_2-\Delta_1}{2} \right)
}{\Gamma(\tilde\Delta)
\Gamma\left(\frac{\Delta+\Delta_1-\Delta_2}{2} \right)
\Gamma\left(\frac{\Delta+\Delta_2-\Delta_1}{2} \right)}.
\end{equation}

The conformal partial wave expansion of the four-point function is its representation in the form  of an integral over conformal partial waves with dimensions in the principal series, $\Delta=h+i\nu$, $\nu \in \mathbb{R}$,
\begin{equation}
\label{21jun1}
A^{\Delta_i}(P_i) = \int_{-\infty}^{\infty}d\nu I^{\Delta_i}_{h+i\nu}\Psi^{\Delta_i}_{h+i\nu}(P_i).
\end{equation}
Substituting  (\ref{2jun12}), one finds
\begin{equation}
\label{21jun2}
\begin{split}
A^{\Delta_i}(P_i) &= \int_{-\infty}^{\infty}d\nu I^{\Delta_i}_{h+i\nu}\left( S_{h-i\nu}^{\Delta_3,\Delta_4} G^{\Delta_i}_{h+i\nu}(P_i)+
S_{h+i\nu}^{\Delta_1,\Delta_2} G^{\Delta_i}_{h-i\nu}(P_i)\right)\\
&\qquad\qquad\qquad\qquad=
\int_{-\infty}^{\infty}d\nu \left( I^{\Delta_i}_{h+i\nu}S_{h-i\nu}^{\Delta_3,\Delta_4}+
 I^{\Delta_i}_{h-i\nu}
S_{h-i\nu}^{\Delta_1,\Delta_2}\right) G^{\Delta_i}_{h+i\nu}(P_i).
\end{split}
\end{equation}
The coefficient function $I$ can always be split into two parts
\begin{equation}
\label{21jun3}
I^{\Delta_i}_{h+i\nu}=I^{\Delta_i;(+)}_{h+i\nu}+I^{\Delta_i;(-)}_{h+i\nu},
\qquad
I^{\Delta_i;(\pm)}_{h+i\nu}S_{h-i\nu}^{\Delta_3,\Delta_4}= \pm I^{\Delta_i;(\pm)}_{h-i\nu}
S_{h-i\nu}^{\Delta_1,\Delta_2}.
\end{equation}
Clearly, $I^{(-)}$ does not contribute to (\ref{21jun2}), so without loss of generality, we can assume that $I^{(-)}$ is vanishing. As a result, we get
\begin{equation}
\label{21jun4}
\begin{split}
A^{\Delta_i}(P_i) = 
\int_{-\infty}^{\infty}d\nu  C^{\Delta_i}_{h+i\nu} G^{\Delta_i}_{h+i\nu}(P_i)
\end{split}
\end{equation}
where
\begin{equation}
\label{21jun4x1}
  C^{\Delta_i}_{h+i\nu} =2I^{\Delta_i}_{h+i\nu}S_{h-i\nu}^{\Delta_3,\Delta_4}.
\end{equation}

Provided that $P_i$ are in a kinematic regime where the OPE is valid, $G_{h+i\nu}^{\Delta_i}$ decays exponentially in the lower half $\nu$-plane, so we can close the contour in that direction. The resulting integral is then evaluated employing the residue theorem. Assuming that singularities of $C$ within the contour occur at
$h+i\nu=\Delta_n$  we find
\begin{equation}
\label{19jun4}
\begin{split}
A^{\Delta_i}(P_i) =-2\pi \sum_n \res_{h+i\nu=\Delta_n} \left( C^{\Delta_i}_{h+i\nu} G^{\Delta_i}_{h+i\nu}(P_i)\right).
\end{split}
\end{equation}
If $h+i\nu =\Delta_n$ is a pole of order $m$, we get
\begin{equation}
\label{19jun4x1}
\begin{split}
&\res_{h+i\nu=\Delta_n} \left( C^{\Delta_i}_{h+i\nu} G^{\Delta_i}_{h+i\nu}(P_i)\right)\\
&\qquad\qquad\qquad= \frac{1}{(m-1)!}\lim_{\Delta\to\Delta_n}\left[\frac{\partial^{m-1}}{\partial\Delta^{m-1}}\left((\Delta-\Delta_n)^m C^{\Delta_i}_{\Delta} G^{\Delta_i}_{\Delta}(P_i)\right) \right]\\
&\qquad\qquad\qquad=\sum_{k=0}^{m-1} \frac{1}{k!(m-1-k)!} \left. \frac{\partial^{m-1-k}}{\partial\Delta^{m-1-k}} \kappa^{\Delta_i}_{\Delta}(\Delta_n,m) \frac{\partial^{k}}{\partial\Delta^{k}}G^{\Delta_i}_{\Delta} (P_i)\right|_{\Delta=\Delta_n},
\end{split}
\end{equation}
where 
\begin{equation}
\label{19jun5}
\kappa^{\Delta_i}_{\Delta} (\Delta_n,m)\equiv (\Delta-\Delta_n)^m C^{\Delta_i}_{\Delta}.
\end{equation}
Therefore, higher order singularities in the expansion (\ref{21jun4}) result into the presence of terms with derivatives of conformal blocks in the conformal block decomposition. Such terms are absent in complete CFT's, however, they do occur for perturbative bulk computations and for CFT's in the large-$N$ expansion. 
We will  denote the coefficients of such a conformal block decomposition  as 
\begin{equation}
\label{21jun5}
A^{\Delta_i}(P_i) = \sum_{n} \sum_{k=0} a^{[k]}_{\Delta_n} \frac{1}{k!}\left.\frac{\partial^k}{\partial \Delta^k} G^{\Delta_i}_{\Delta} (P_i)\right|_{\Delta=\Delta_n}.
\end{equation}
In view of the connection with the boundary theory, it is more conventional to express the coefficients of derivatives of conformal blocks in terms of anomalous dimensions, as we review in section \ref{sec:largeN}.

\subsection{Four-point functions}
\label{sec:4pt}
Below we illustrate how the split representation allows one to find the conformal partial wave expansions for simplest tree-level four-point Witten diagrams. 

The four-point Witten diagram for $g_4 \phi^4$ interaction is defined by
\begin{equation}
\label{2jun17}
\begin{split}
A^{\Delta_i}(P_i)\equiv g_4 \int dX \Pi_{\Delta_1}(X,P_1)\Pi_{\Delta_2}(X,P_2)\Pi_{\Delta_3}(X,P_3)\Pi_{\Delta_4}(X,P_4).
\end{split}
\end{equation}
Its evaluation proceeds as follows. First, one introduces an  additional trivial bulk integration using a delta-function
\begin{equation}
\label{2jun17x1}
\begin{split}
A^{\Delta_i}(P_i)
=g_4 \int dX_1\int dX_2 \Pi_{\Delta_1}(X_1,P_1)\Pi_{\Delta_2}(X_1,P_2)\delta(X_1,X_2)\Pi_{\Delta_3}(X_2,P_3)\Pi_{\Delta_4}(X_2,P_4).
\end{split}
\end{equation}
Next, rewriting the delta-function as in (\ref{2jun5}) and substituting  $\Omega_\nu$ in the split form (\ref{2jun4}), one finds
\begin{equation}
\label{2jun17x2}
\begin{split}
A^{\Delta_i}(P_i)
&=g_4 \int dX_1\int dX_2 \int d\nu \frac{\nu^2}{\pi}\int dP_0
\Pi_{h+i\nu}(X_1,P_0) \Pi_{h-i\nu}(X_2,P_0)\\
&\qquad \qquad \qquad\qquad\qquad\Pi_{\Delta_1}(X_1,P_1)\Pi_{\Delta_2}(X_1,P_2)
\Pi_{\Delta_3}(X_2,P_3)\Pi_{\Delta_4}(X_2,P_4)\\
&=g_4\int_{-\infty}^\infty d\nu \frac{\nu^2}{\pi}b(\Delta_1,\Delta_2,h+i\nu)b(\Delta_3,\Delta_4,h-i\nu) \Psi_{h+i\nu}^{\Delta_i}.
\end{split}
\end{equation}
Bulk integrals in (\ref{2jun17x2}) are evaluated with the help of (\ref{2jun9x1}). This brings the amplitude to the form of the conformal partial wave expansion. Finally, expressing conformal partial waves in terms of conformal blocks we obtain
\begin{equation}
\label{2jun18}
\begin{split}
A^{\Delta_i}(P_i)&=
2 g_4\int_{-\infty}^\infty d\nu \frac{\nu^2}{\pi}S_{h-i\nu}^{\Delta_3,\Delta_4}b(\Delta_1,\Delta_2,h+i\nu)b(\Delta_3,\Delta_4,h-i\nu) G_{h+i\nu}^{\Delta_i}\\
&= g_4 \int_{-\infty}^\infty d\nu B_{h+i\nu}^{\Delta_i}G^{\Delta_i}_{h+i\nu},
\end{split}
\end{equation}
where
\begin{equation}
\label{2jun18x1}
\begin{split}
B_{h+i\nu}^{\Delta_i}&=\frac{\pi^{h-1}}{8} \frac{{\cal C}_{\Delta_1}{\cal C}_{\Delta_2}{\cal C}_{\Delta_3}{\cal C}_{\Delta_4}}{\Gamma(\Delta_1)\Gamma(\Delta_2)\Gamma(\Delta_3)\Gamma(\Delta_4) \Gamma(i\nu)\Gamma(h+i\nu)}\\
&\qquad\quad \Gamma\left(\tfrac{\Delta_1+\Delta_2-h+i\nu}{2} \right)
\Gamma\left(\tfrac{\Delta_1+\Delta_2-h-i\nu}{2} \right)
\Gamma\left(\tfrac{\Delta_3+\Delta_4-h+i\nu}{2} \right)
\Gamma\left(\tfrac{\Delta_3+\Delta_4-h-i\nu}{2} \right)\\
&\qquad\quad\Gamma\left(\tfrac{h+i\nu+\Delta_1-\Delta_2}{2} \right)
\Gamma\left(\tfrac{h+i\nu+\Delta_2-\Delta_1}{2} \right)
\Gamma\left(\tfrac{h+i\nu+\Delta_3-\Delta_4}{2} \right)
\Gamma\left(\tfrac{h+i\nu+\Delta_4-\Delta_3}{2} \right).
\end{split}
\end{equation}

Analogously, one can find the direct channel conformal partial wave expansion for the exchange
\begin{equation}
\label{21jun6}
\begin{split}
A^{\Delta_i}(P_i)
&=g_3g'_3 \int dX_1\int dX_2 \Pi_{\Delta_1}(X_1,P_1)\Pi_{\Delta_2}(X_1,P_2)\\
&\qquad\qquad\qquad\qquad\qquad\qquad\qquad\Pi_\Delta(X_1,X_2)\Pi_{\Delta_3}(X_2,P_3)\Pi_{\Delta_4}(X_2,P_4),
\end{split}
\end{equation}
where $g_3$ and $g_3'$ are two cubic bulk coupling constants. 
The  only difference between (\ref{21jun6}) and (\ref{2jun17x1}) is that the delta-function for the contact interaction is replaced with the bulk-to-bulk propagator for the exchange. Proceeding in the same way as before we find
\begin{equation}
\label{21jun7}
\begin{split}
A^{\Delta_i}(P_i)
=g_3g'_3\int_{-\infty}^\infty d\nu \frac{1}{ \nu^2 + (\Delta-h)^2} B_{h+i\nu}^{\Delta_i}  G_{h+i\nu}^{\Delta_i}.
\end{split}
\end{equation}

Using the same procedure, one can evaluate any contact four-point amplitude \cite{Bekaert:2015tva} as well as any exchange in the direct channel \cite{Costa:2014kfa,Bekaert:2014cea}. The result has the form of the conformal partial wave expansion, in which the coefficient function can be conveniently factorized
\begin{equation}
\label{25jun2}
\begin{split}
A^{\Delta_i}(P_i) = 
\int_{-\infty}^{\infty}d\nu  A^{\Delta_i}_{h+i\nu} B^{\Delta_i}_{h+i\nu} G^{\Delta_i}_{h+i\nu}(P_i).
\end{split}
\end{equation}
Here $B$ is the universal kinematic factor (\ref{2jun18x1}), while $A$ is  characteristic of a particular bulk process: $A$ turns out to be polynomial in $\nu$ for contact diagrams and has poles for the direct channel exchanges. This motivates why amplitudes with regular $A$ are considered regular, while the pole part of $A$ and the associated contributions to the conformal block decomposition are considered to be the singular part of the amplitude\footnote{Nowadays, it became conventional to define the singular part of the correlator as its double discontinuity in the coordinate representation on the boundary \cite{Caron-Huot:2017vep,Aprile:2017bgs,Alday:2017vkk,Aprile:2017qoy,Aharony:2018npf,Alday:2018pdi,Alday:2018kkw,Alday:2019qrf,Alday:2019clp}. This definition of the singular part is consistent with the one given in the text in the sense that only the singularities of $A$ result into terms in the conformal block decomposition with non-vanishing double discontinuity. It is  worth mentioning that the singular part can also be defined in the Mellin space as the pole part of the Mellin amplitude \cite{Penedones:2010ue,Fitzpatrick:2011hu,Fitzpatrick:2011dm}. The latter definition differs from previous ones in the regard that it is channel-independent and captures singularities in all channels simultaneously. It was demonstrated that these definitions of the singular part reproduce singularities of flat space amplitudes in  the flat space limit \cite{Penedones:2010ue,Fitzpatrick:2011hu,Fitzpatrick:2011dm,Alday:2017vkk}.}.

To obtain the conformal block decomposition from (\ref{25jun2}), we start by closing the integration contour in the lower half-$\nu$ plane. When identifying the singularities of the integrand that are located inside the contour it is important to keep in mind that  (\ref{25jun2}) can be applied literally only when the external dimensions $\Delta_i$ are in the principal series. For $\Delta_i$ away from the principal series the integral should be defined by analytic continuation to the required values of $\Delta_i$. This implies that if while changing $\Delta_i$ some poles of $C$ cross the real $\nu$ line, the integration contour should be properly indented, so that all singularities of $C$ remain on the same side of the contour. 

To start,  consider singularities generated by $B$ (\ref{2jun18x1}). It contains a product of eight gamma functions, which generate eight series of poles. However, for $\Delta_i$ in the principal series, only the poles from 
\begin{equation}
\label{25jun1}
\Gamma\left(\tfrac{\Delta_1+\Delta_2-h-i\nu}{2} \right) \qquad \text{and}
\qquad \Gamma\left(\tfrac{\Delta_3+\Delta_4-h-i\nu}{2} \right)
\end{equation}
appear in the lower half complex $\nu$-plane. Similarly, only these singularities should be considered when reducing (\ref{25jun2}) to the conformal block decomposition for $\Delta_i$ in the case of interest --- that is for all $\Delta_i$ real and satisfying the unitarity bound.

Depending on the values of $\Delta_i$, we need to consider two situations, for which the analysis is different, especially on the CFT side. First, we consider the case
\begin{equation}
\label{25jun3}
\frac{1}{2}(\Delta_1+\Delta_2-\Delta_3-\Delta_4) \notin \mathbb{Z}.
\end{equation}
For these values of $\Delta_i$ the two series of poles from (\ref{25jun1}) do not overlap and produce conformal blocks of dimensions 
\begin{equation}
\label{25jun4}
\Delta_1+\Delta_2+2n, \qquad \Delta_3+\Delta_4+2n, \qquad n\in \mathbb{Z}, \quad n\ge 0
\end{equation}
in the conformal block decomposition. Otherwise, that is for
\begin{equation}
\label{25jun4x1}
\frac{1}{2}(\Delta_1+\Delta_2-\Delta_3-\Delta_4) \in \mathbb{Z},
\end{equation}
 two series of poles overlap and produce double poles. This situation occurs, for example, when we are dealing with the amplitude of four identical scalar fields.  Applying the residue theorem at these singularities one finds that the conformal block decomposition of the amplitude contains not only conformal blocks, but also their first derivatives.

In addition to the singularities of $B$, producing regular terms in the amplitude,  one should take into account singular contributions generated from poles of $A$. These are evaluated in a similar way. For example, for the exchange (\ref{21jun7}) this results in a conformal block of dimension $\Delta$. As for the regular contributions, there can be various  degeneracies in  locations of poles of $A$ with itself as well as with poles of $B$. These cases are analyzed analogously.

\section{One-loop amplitude from tree-level amplitudes}
\label{sec:loopfromtree}

In this section we consider a one-loop four-point amplitude with a non-trivial two-particle cut in a given channel and show how the methods reviewed in the previous section combined with the bubble integral formula can be used to obtain its conformal partial wave expansion in terms of the conformal partial wave expansions for tree-level subdiagrams, resulting from a cut of the original diagram. 

To start, we use the split representation for two propagators, that will be cut in the following. We will label by $Q_i$ the external lines of the amplitude and by $P_i$ the two extra boundary points introduces by the split representation. Then, the amplitude can be factorized into the amplitudes for subdiagrams as follows
\begin{equation}
\label{6jun2}
\begin{split}
A_{\rm O}(Q_i)&=\int d\nu_1d\nu_2 dP_1 dP_2 \frac{\nu_1^2}{\pi}\frac{\nu_2^2}{\pi}\frac{1}{\nu_1^2+(\Delta_1-h)^2}\frac{1}{\nu_2^2+(\Delta_2-h)^2}\\
 &\qquad A_{\rm L}^{\Delta_1^{\rm e},\Delta_2^{\rm e},h+i\nu_1,h+i\nu_2} (Q_1,Q_2;P_1,P_2)\cdot
A_{\rm R}^{\Delta_3^{\rm e},\Delta_4^{\rm e},h-i\nu_1,h-i\nu_2} (Q_3,Q_4;P_1,P_2).
\end{split}
\end{equation}
Here $A_{\rm O}$ is the original loop amplitude, 
$A_{\rm L}$ and $A_{\rm R}$ are the amplitudes for its left and right subdiagrams and 
the superscript ${\rm e}$ for dimensions refers to external lines. It is worth emphasizing that both tree-level amplitudes have two out of four external lines off-shell.

Next, we substitute the conformal partial wave expansion for the left subdiagram
\begin{equation}
\begin{split}
\label{6jun3}
&A_{\rm L}^{\Delta_1^{\rm e},\Delta_2^{\rm e},h+i\nu_1,h+i\nu_2} (Q_1,Q_2;P_1,P_2)\\
&\qquad\qquad\qquad\qquad=
\int d\nu_L I^{\Delta_1^{\rm e},\Delta_2^{\rm e},h+i\nu_1,h+i\nu_2}_{{\rm L}|h+i\nu_{\rm L}}
\Psi_{h+i\nu_{\rm L}}^{\Delta_1^{\rm e},\Delta_2^{\rm e},h+i\nu_1,h+i\nu_2}(Q_1,Q_2;P_1,P_2)
\end{split}
\end{equation}
and the analogous one for the right subdiagram into (\ref{6jun2}). As a result, we obtain
\begin{equation}
\label{6jun4}
\begin{split}
A_{\rm O}(Q_i)=&\int d\nu_1 d\nu_2 \frac{\nu_1^2\nu_2^2}{\pi^2}\frac{1}{\nu_1^2+(\Delta_1-h)^2}\frac{1}{\nu_2^2+(\Delta_2-h)^2}
\\
&\int d\nu_{\rm L} d\nu_{\rm R} I^{\Delta_1^{\rm e},\Delta_2^{\rm e},h+i\nu_1,h+i\nu_2}_{L|h+i\nu_{\rm L}}
I^{\Delta_3^{\rm e},\Delta_4^{\rm e},h-i\nu_1,h-i\nu_2}_{{\rm R}|h+i\nu_{\rm R}}\\
&\int dP_1dP_2 
\Psi_{h+i\nu_{\rm L}}^{\Delta_1^{\rm e},\Delta_2^{\rm e},h+i\nu_1,h+i\nu_2}(Q_1,Q_2;P_1,P_2)
\Psi_{h+i\nu_{\rm R}}^{\Delta_3^{\rm e},\Delta_4^{\rm e},h-i\nu_1,h-i\nu_2}(Q_3,Q_4;P_1,P_2).
\end{split}
\end{equation}

\subsection{Integrated product of conformal partial waves}
The last line of (\ref{6jun4}) has the form of a product of two conformal partial waves integrated over two common points
\begin{equation}
\label{16jul1}
{\cal I}\equiv 
\int dP_1dP_2 
\Psi_{h+i\nu_L}^{\Delta_1^{\rm e},\Delta_2^{\rm e},h+i\nu_1,h+i\nu_2}(Q_1,Q_2;P_1,P_2)
\Psi_{h+i\nu_R}^{\Delta_3^{\rm e},\Delta_4^{\rm e},h-i\nu_1,h-i\nu_2}(Q_3,Q_4;P_1,P_2).
\end{equation}
 It can be simplified as follows. 
 
 First, we use the definition of conformal  partial waves (\ref{2jun11}), introducing two additional integrals over $P_{\rm L}$ and $P_{\rm R}$
\begin{equation}
\label{6jun4x1}
\begin{split}
{\cal I}
&=
\int dP_1dP_2 dP_{\rm L} dP_{\rm R}
[ {\cal O}_{\Delta^{\rm e}_1}(Q_1) {\cal O}_{\Delta^{\rm e}_2}(Q_2) {\cal O}_{h+i\nu_{\rm L}}(P_{\rm L}) ]
[ {\cal O}_{h-i\nu_{\rm L}}(P_{\rm L}) {\cal O}_{h+i\nu_1}(P_1) {\cal O}_{h+i\nu_2}(P_2) ]\\
&\qquad\qquad\qquad\qquad[ {\cal O}_{h-i\nu_1}(P_1) {\cal O}_{h-i\nu_2}(P_2) {\cal O}_{h+i\nu_{\rm R}}(P_{\rm R}) ]
[ {\cal O}_{h-i\nu_{\rm R}}(P_{\rm R}) {\cal O}_{\Delta^{\rm e}_3}(Q_3) {\cal O}_{\Delta^{\rm e}_4}(Q_4) ].
\end{split}
\end{equation}
Then we find that integrals over $P_1$ and $P_2$ in (\ref{6jun4x1}) can be evaluated using the bubble integral formula  \cite{Dobrev:1976vr,Karateev:2018oml}
\begin{equation}
\label{6jun1}
\begin{split}
&\int dP_1 dP_2 [ {\cal O}_{\tilde\Delta_1}(P_1)  {\cal O}_{\tilde\Delta_2}(P_2) {\cal O}_{\Delta_0}(P_0)]
[ {\cal O}_{\Delta_1}(P_1)  {\cal O}_{\Delta_2}(P_2) {\cal O}_{\tilde\Delta_3}(P_3)]\\
&\qquad\qquad=4\pi \frac{\pi^{3h}\Gamma(\Delta_0-h)\Gamma(h-\Delta_0)}{\Gamma(h)\Gamma(\Delta_0)\Gamma(d-\Delta_0)}
\left( \delta(\nu_0-\nu_3)\delta(P_0,P_3)
\phantom{\frac{\Gamma\left(\frac{\tilde\Delta_0}{2}\right)}{\Gamma\left(\frac{\Delta_0}{2}\right)}}\right.
\\
&\left.\qquad\qquad\qquad\qquad+\delta(\nu_0+\nu_3)\frac{\Gamma\left(\frac{\tilde\Delta_0+\Delta_2-\Delta_1}{2} \right)\Gamma\left(\frac{\tilde\Delta_0+\Delta_1-\Delta_2}{2} \right)}{\Gamma\left(\frac{\Delta_0+\Delta_2-\Delta_1}{2} \right)\Gamma\left(\frac{\Delta_0+\Delta_1-\Delta_2}{2} \right)}
\frac{\Gamma(\Delta_0)}{\Gamma(h-\Delta_0)}\frac{1}{\pi^h}\frac{1}{P^{\Delta_0}_{03}}\right),
\end{split}
\end{equation}
where $\nu_0=-i (\Delta_0-h)$ and $\nu_3 = -i (\Delta_3-h)$. Employing it inside (\ref{6jun4x1}), we obtain
\begin{equation}
\label{6jun5}
\begin{split}
{\cal I}
&=\int dP_{\rm L} dP_{\rm R}
[ {\cal O}_{\Delta_1^{\rm e}}(Q_1) {\cal O}_{\Delta_2^{\rm e}}(Q_2) {\cal O}_{h+i\nu_{\rm L}}(P_{\rm L}) ]
[ {\cal O}_{h-i\nu_{\rm R}}(P_{\rm R}) {\cal O}_{\Delta_3^{\rm e}}(Q_3) {\cal O}_{\Delta_4^{\rm e}}(Q_4)  ]\\
&\qquad
4\pi \frac{\pi^{3h}\Gamma(i\nu_{\rm R})\Gamma(-i \nu_{\rm R})}{\Gamma(h)\Gamma(h+i\nu_{\rm R})\Gamma(h-i\nu_{\rm R})}
\left( \delta(\nu_{\rm R}-\nu_{\rm L})\delta(P_{\rm R},P_{\rm L})
\phantom{\frac{\Gamma\left(\frac{h}{2}\right)}{\Gamma\left(\frac{h}{2}\right)}}\right.
\\
&\left.\qquad\quad+\delta(\nu_{\rm R}+\nu_{\rm L})\frac{\Gamma\left(\frac{h-i\nu_{\rm R}+i\nu_2-i\nu_1}{2} \right)\Gamma\left(\frac{h-i\nu_{\rm R}+i\nu_1-i\nu_2}{2} \right)}{\Gamma\left(\frac{h+i\nu_{\rm R}+i\nu_2-i\nu_1}{2} \right)\Gamma\left(\frac{h+i\nu_{\rm R}+i\nu_1-i\nu_2}{2} \right)}
\frac{\Gamma(h+i\nu_{\rm R})}{\Gamma(-i\nu_{\rm R})}\frac{1}{\pi^h} \frac{1}{P_{{\rm L}{\rm R}}^{h+i\nu_{\rm R}}}\right).
\end{split}
\end{equation}

The first term inside brackets in (\ref{6jun5}) has $\delta(P_{\rm R},P_{\rm L})$, which makes $P_{\rm R}$ integration trivial. The remaining $P_{\rm L}$ integral just gives the definition of the conformal partial wave. The second term produces the identical contribution. To see that one should first evaluate the integral over $P_{\rm R}$ with the Symanzik star formula \cite{Symanzik:1972wj,Mack:2009mi,Penedones:2010ue,Paulos:2011ie} and take into account the fact that ${\cal I}$ is supposed to be integrated over $\nu_{\rm L}$ and $\nu_{\rm R}$  against $I_{\rm L}$ and $I_{\rm R}$ that satisfy the symmetry property discussed below (\ref{21jun3}). This is a straightforward computation and we leave it to the reader.
Adding both terms, we find
\begin{equation}
\label{6jun7}
\begin{split}
{\cal I}
=
8\pi \frac{\pi^{3h}\Gamma(i\nu_{\rm L})\Gamma(-i \nu_{\rm L})}{\Gamma(h)\Gamma(h+i\nu_{\rm L})\Gamma(h-i\nu_{\rm L})}
\delta(\nu_{\rm R}-\nu_{\rm L})\Psi_{h+i\nu_{\rm L}}^{\Delta_i^{\rm e}}(Q_i).
\end{split}
\end{equation}

\subsection{Collecting the results}
Now we come back to the loop amplitude (\ref{6jun4}) we were computing. With (\ref{6jun7}) we find

\begin{equation}
\label{6jun11}
\begin{split}
A_{\rm O}(Q_i)&=\int d\nu_{\rm L}  8\pi \frac{\pi^{3h}\Gamma(i\nu_{\rm L})\Gamma(-i \nu_{\rm L})}{\Gamma(h)\Gamma(h+i\nu_{\rm L})\Gamma(h-i\nu_{\rm L})}\\
&\qquad\qquad\qquad\int d\nu_1 d\nu_2 \frac{\nu_1^2\nu_2^2}{\pi^2}\frac{1}{\nu_1^2+(\Delta_1-h)^2}\frac{1}{\nu_2^2+(\Delta_2-h)^2}\\
 &\qquad\qquad\qquad\qquad\qquad\qquad I^{\Delta_1^{\rm e},\Delta_2^{\rm e},h+i\nu_1,h+i\nu_2}_{{\rm L}| h+i\nu_{\rm L}}
I^{\Delta_3^{\rm e},\Delta_4^{\rm e},h-i\nu_1,h-i\nu_2}_{{\rm R}|h+i\nu_{\rm L}}\Psi_{\Delta_{\rm L}}^{\Delta_i^{\rm e}}(Q_i).
\end{split}
\end{equation}
This formula gives the conformal partial wave expansion for the one-loop amplitude in terms of tree-level data.
The coefficient function of the conformal partial wave expansion for the one-loop amplitude is then given by
\begin{equation}
\label{6jun12}
\begin{split}
&I_{\rm O}(\nu;\Delta_i^{\rm e})=
 \frac{8\pi^{3h-1}\Gamma(i\nu)\Gamma(-i \nu)}{\Gamma(h)\Gamma(h+i\nu)\Gamma(h-i\nu)}\\
&\quad\int\int d\nu_1 d\nu_2 \frac{\nu_1^2}{\nu_1^2+(\Delta_1-h)^2}\frac{{\nu_2^2}}{\nu_2^2+(\Delta_2-h)^2}
I^{\Delta_1^{\rm e},\Delta_2^{\rm e},h+i\nu_1,h+i\nu_2}_{{\rm L}| h+i\nu}
I^{\Delta_3^{\rm e},\Delta_4^{\rm e},h-i\nu_1,h-i\nu_2}_{{\rm R}|h+i\nu}.
\end{split}
\end{equation}

\section{Singularities of one-loop amplitudes}
\label{sec:sing}

In the previous section we expressed a one-loop amplitude with a non-trivial double-particle cut in terms of tree-level subamplitudes. The resulting formula (\ref{6jun11}), (\ref{6jun12}) is exact in the sense that no terms were omitted when it was derived.
 At the same time, tree-level amplitudes involved in this formula feature off-shell fields on external lines. Below we will show that, 
 as in flat space, 
  the singular part of the one-loop amplitude associated with a given double-particle cut is defined purely in terms of tree-level diagrams with all external lines being on-shell. We will then recast the result into the form suitable for comparison with the CFT.
At a more technical level, in (\ref{6jun12}) the coefficient function for the one-loop amplitude $I_{\rm O}$ is expressed as a double integral of a weighted product of coefficient functions of tree-level amplitudes $I_{\rm L}$ and $I_{\rm R}$. Instead of evaluating the integrals exactly, one can study their analytic structure employing the standard techniques, see e.g. \cite{smatrix,Yuan:2018qva}. 

\subsection{Locations of singularities}
As a first step, we need to understand the analytic structure of the integrand in variables $\nu_1$ and $\nu_2$. 
Besides the explicit poles coming from propagators, the right hand side of (\ref{6jun12})
 also has two coefficient functions $I_{\rm L}$ and $I_{\rm R}$ for tree-level diagrams. In the examples, considered in section \ref{sec:4pt} the dependence of the latter on $\nu_1$ and $\nu_2$ was given by
\begin{equation}
\label{27jun1}
\begin{split}
 I_{{\rm L}|h+i\nu}^{\Delta_1^{\rm e},\Delta_2^{\rm e},h+i\nu_1,h+i\nu_2}& \propto b(h-i\nu,h+i\nu_1,h+i\nu_2), \\
 I_{{\rm R}|h+i\nu}^{\Delta_3^{\rm e},\Delta_4^{\rm e},h-i\nu_1,h-i\nu_2} &\propto b(h+i\nu,h-i\nu_1,h-i\nu_2).
 \end{split}
\end{equation}
It is not hard to see, that (\ref{27jun1}) correctly captures singularities of the coefficients functions in $\nu_1$ and $\nu_2$ in general.  Indeed, the type of a bulk processes associated with $A_{\rm L}$ and $A_{\rm R}$ only affects the $\nu$ dependence. The only way how the $\nu_1$ and $\nu_2$ dependence can be changed is if the bulk vertices involve derivatives of external lines $P_1$ and $P_2$. Such derivatives, however, generate only polynomial contributions in $\nu_1$ and $\nu_2$. For example, employing that 
\begin{equation}
\label{27jun2}
\nabla^2\Pi_{h+i\nu_1} = -(h^2+\nu_1^2)\Pi_{h+i\nu_1},
\end{equation}
we find that 
a d'Alembertian acting on the external line $P_1$ of $A_{\rm L}$ produces an additional factor of $-(h^2+\nu_1^2)$ for the coefficient function $I_{\rm L}$, which does not bring any new singularities. Thus, unless we are dealing with a non-local theory, in which such polynomial terms may sum up to a singularity, all singularities of the coefficient functions $I_{\rm L}$ and $I_{\rm R}$ are produced by the $b$ factors as stated in (\ref{27jun1}). 

To summarize, in total, the integrand in (\ref{6jun12}) has the following analytic structure: it has two pairs of poles generated by the two propagators
\begin{equation}
\label{27jun3}
\frac{1}{\nu_1^2+(\Delta_1-h)^2}, \qquad \frac{1}{\nu_2^2+(\Delta_2-h)^2}
\end{equation}
and series of poles generated by the coefficient functions of tree-level diagrams
\begin{equation}
\label{27jun5}
b(h-i\nu,h+i\nu_1,h+i\nu_2) \; b(h+i\nu,h-i\nu_1,h-i\nu_2).
\end{equation}

With the analytic structure of the integrand clarified, we proceed with the analytic structure of the integral itself. To this end, we use the standard argument, which goes as follows.  First, one notices that for real $\nu$ all poles (\ref{27jun3})-(\ref{27jun5}) are away from the real axis, where the $\nu_1$ and $\nu_2$ integration contours are located.  Therefore, we can conclude that the integral is regular for real $\nu$\footnote{This integral can be divergent and then it has to be regularized by subtracting counterterms. The counterterms are, however, regular, and do not affect the analytic structure of the integral.}.
Next, one considers the analytic continuation of the integral to the complex $\nu$ plane. When $\nu$ moves away from the real axis, poles (\ref{27jun3})-(\ref{27jun5}) also move and the integration contours should be deformed so that singularities do not cross them. 
The integral remains analytic in $\nu$ unless the integration contours get pinched by the singularities, which prevents their further deformations. In other words, all singularities of the integral can be found by studying the configurations in which the integration contours get pinched by the singularities of the integrand.

For (\ref{6jun12}) the analysis of the analytic structure is straightforward, but somewhat cumbersome, due to the presence of several series of poles and zeros in (\ref{27jun5}) as well as due to extra zeros in the explicit prefactor in (\ref{6jun12}). Luckily, the integral with the same analytic structure was analyzed in  \cite{Penedones:2010ue}, where the Mellin amplitude for a bubble diagram was studied. We will not repeat this analysis in detail here and just quote the end result\footnote{We also illustrate the key features of this analysis with a toy example in Appendix \ref{App:Ex}.}.
Namely, one finds that $I_{\rm O}$ has singularities at
\begin{equation}
\label{27jun6}
h\pm i\nu = \Delta_1+\Delta_2+2n, \qquad n\in \mathbb{Z},\; n\ge 0. 
\end{equation}
These, for example, occur when the contour is trapped between three singularities
\begin{equation}
\label{27jun7}
\{ \nu_1=-i (\Delta_1-h), \qquad \nu_2=-i(\Delta_2-h),\qquad h+i\nu_1+i\nu_2-i\nu = -2n,\; n\in \mathbb{Z}, \; n\ge 0 \},
\end{equation}
which gives a series (\ref{27jun6}) with a plus sign on the left hand side.
Other seven series of pinching configurations are related to (\ref{27jun7}) by the symmetry with respect to three independent reflections $\nu\to -\nu$, $\nu_1\to -\nu_1$ and $\nu_2\to -\nu_2$.

\subsection{Residues}
Once locations of poles of the integral are identified, we can proceed by specifying their residues in the standard way. When the pinching configuration occurs, one can always split the integration contour into two parts, so that the first part is free of any pinches, while the second part consists of an infinitesimal contour encircling one of the singularities. Then, the singular part of the  integral associated with a given pinch configuration remains the same if we replace the original contour with its second part. The latter, in turn, can be evaluated by the residue theorem. 

For the pinch configuration (\ref{27jun7}) the singularity of the integral can be captured by replacing the $\nu_i$ integration contours along the real axes with infinitesimal circular contours around $\nu_1=-i (\Delta_1-h)$ and $\nu_2=-i(\Delta_2-h)$. Evaluation of the latter integrals reduces to the evaluation of residues at these poles. Taking into account the symmetry of the integral $\nu_1\to-\nu_1$ and $\nu_2\to -\nu_2$, we get an extra factor of four. Summing up, we find 
\begin{equation}
\label{27jun8}
\begin{split}
I_{{\rm O}|h+i\nu}^{\Delta_i^{\rm e}}=
32 \frac{\pi^{3h+1}\Gamma(i\nu)\Gamma(-i\nu)}{\Gamma(h)\Gamma(h+i\nu)\Gamma(h-i\nu)}
(h-\Delta_1)(h-\Delta_2)
 I_{{\rm L}| h+i\nu}^{\Delta_1^{\rm e},\Delta_2^{\rm e},\Delta_1,\Delta_2}
I_{{\rm R}|h+i\nu }^{\Delta_3^{\rm e},\Delta_4^{\rm e},\tilde\Delta_1,\tilde\Delta_2}\\
+\text{ less singular terms}.
\end{split}
\end{equation}

Let us be more precise with what is captured by the explicit term in (\ref{27jun8}) and what we mean by ''less singular terms''. 
The argument presented above implies that the integral on the right hand side of (\ref{6jun12}) with $I_{\rm L}$ and $I_{\rm R}$ depending on $\nu_i$ as in (\ref{27jun1}) produces simple poles in $\nu$ at double-trace\footnote{Here ''double-trace'' refers to double-trace operators on the CFT side, associated with these singularities. This terminology is standard in the AdS/CFT literature and will be further explained in section \ref{sec:largeN}, in which we discuss the CFT dual picture. It is worth stressing that in the present setup we encounter two types of double-trace operators -- those built of pairs of operators on external lines of the Witten diagram and those built of operators running in the loop. The associated contributions play different roles in the bulk analysis, see section \ref{sec:singsum}.} locations (\ref{27jun6}), moreover, residues of the original integral and the integral over a deformed contour at these locations are the same. In the following, we will encounter situations in which $I_{\rm L}$ and $I_{\rm R}$ themselves have poles in $\nu$ at (\ref{27jun6}). Clearly, in this case the $\nu_1$ and $\nu_2$ integration will increase the total order of the pole of $I_{\rm L}$ and $I_{\rm R}$  by one. Then the explicit term in (\ref{27jun8}) can be used to compute reliably the coefficients of the highest order poles of $I_{\rm O}$ at double-trace locations, while ''less singular terms'' refers to all other contributions to $I_{\rm O}$, that do not affect the leading order double-trace singularity. It is worth stressing here that in addition to  double-trace singularities occurring at (\ref{27jun6}), the explicit term in (\ref{27jun8}) also has singularities at various shadow double-trace locations. These should be ignored. This issue is illustrated in Appendix \ref{App:Ex} and will be further discussed below.

\subsection{Translation to the CFT language}

We have just derived formula (\ref{27jun8}), which relates the conformal partial wave coefficient functions for the singular part of the one-loop amplitude and for its tree level subdiagrams. Before we will be able to rewrite it in the form suitable for making the connection with the large-$N$ expansion on the boundary side, we need to make a couple of straightforward manipulations. 

 First, we want to express $I_{\rm R}$ featuring shadow dimensions $\tilde\Delta_1$ and $\tilde\Delta_2$ on external lines in terms of an analogous coefficient function for physical dimensions. These are related by a properly normalized shadow transform, which, as it is not hard to see, gives
\begin{equation}
\label{7jun1}
\begin{split}
I_{{\rm R}|h+i\nu}^{\Delta_3^{\rm e},\Delta_4^{\rm e},\tilde\Delta_1,\tilde\Delta_2}
=\frac{b(\tilde\Delta_1,\tilde\Delta_2,h+i\nu)}{b(\Delta_1,\Delta_2,h+i\nu)}
I_{{\rm R}|h+i\nu}^{\Delta_3^{\rm e},\Delta_4^{\rm e},\Delta_1,\Delta_2}.
\end{split}
\end{equation}

Next, we would like to  account for different normalizations on the AdS and the CFT sides as discussed  below (\ref{2jun6x1}). 
We will use an extra bar to indicate quantities given in the CFT normalization. 

Then, we convert coefficient functions of the conformal partial wave expansions to the coefficient functions of conformal blocks (\ref{21jun4}). Combining everything together, we find
\begin{equation}
\label{7jun7}
\begin{split}
\bar C_{{\rm O}|h+i\nu}^{\Delta_i^e}&=
16 \frac{\pi^{3h+1}\Gamma(i\nu)\Gamma(-i\nu)}{\Gamma(h)\Gamma(h+i\nu)\Gamma(h-i\nu)}
\frac{{\cal C}_{\Delta_1}{\cal C}_{\Delta_2}}{S_{h-i\nu}^{\Delta_1,\Delta_2}}
(h-\Delta_1)(h-\Delta_2)\\
&\qquad\quad\frac{b(\tilde\Delta_1,\tilde\Delta_2,h+i\nu)}{b(\Delta_1,\Delta_2,h+i\nu)}
 \bar C_{{\rm L}|h+i\nu}^{\Delta_1^{\rm e},\Delta_2^{\rm e},\Delta_1,\Delta_2}
\bar C_{{\rm R}| h+i\nu}^{\Delta_3^{\rm e},\Delta_4^{\rm e},\Delta_1,\Delta_2}+\text{ less singular terms}\\
&=4 \pi
 \frac{\Gamma(i\nu)\Gamma(\Delta_1)\Gamma(\Delta_2)}{\Gamma(h)\Gamma(h-i\nu)\Gamma(h-\Delta_1)\Gamma(h-\Delta_2)}\\
&\qquad \frac{\Gamma\left(\frac{h-i\nu+\Delta_1-\Delta_2}{2}\right)
\Gamma\left(\frac{h-i\nu+\Delta_2-\Delta_1}{2}\right)
\Gamma\left(\frac{\tilde\Delta_1+\tilde\Delta_2-h+i\nu}{2}\right)
\Gamma\left(\frac{\tilde\Delta_1+\tilde\Delta_2-h-i\nu}{2}\right)
}{\Gamma\left(\frac{h+i\nu+\Delta_1-\Delta_2}{2}\right)
\Gamma\left(\frac{h+i\nu+\Delta_2-\Delta_1}{2}\right)
\Gamma\left(\frac{\Delta_1+\Delta_2-h+i\nu}{2}\right)
\Gamma\left(\frac{\Delta_1+\Delta_2-h-i\nu}{2}\right)}
\\
& \qquad\qquad\qquad\qquad\qquad\quad
\bar C_{{\rm L}|h+i\nu}^{\Delta_1^{\rm e},\Delta_2^{\rm e},\Delta_1,\Delta_2}
\bar C_{{\rm R}| h+i\nu}^{\Delta_3^{\rm e},\Delta_4^{\rm e},\Delta_1,\Delta_2}+\text{ less singular terms}.
\end{split}
\end{equation}

\subsubsection{Conformal block coefficients}

Finally, we would like to rewrite (\ref{7jun7}) as a relation between the coefficients of the conformal block decompositions.
In the same way as in  (\ref{27jun8}), one can argue that in (\ref{7jun7}) the explicit term correctly captures the leading order singularities at locations (\ref{27jun6}).
Furthermore, only singularities in the lower half $\nu$-plane are relevant for the conformal block decomposition. Then, using (\ref{19jun4})-(\ref{21jun5}), we can find the coefficients of the highest derivative terms in the conformal block decomposition for the loop diagram in terms of analogous coefficients for tree-level subdiagrams. 

Let us be more explicit. Consider a case in which the left tree diagram has the pole of order $m_{\rm L}$ at $\Delta_n\equiv\Delta_1+\Delta_2+2n$, while the right one has the order $m_{\rm R}$ singularity. 
 Then, given that the prefactor on the right hand side of (\ref{7jun7}) has zeros at double-trace locations, the order of the singularity for $\bar C_{\rm O}$ is $m_L+m_R-1$.
In this case, the highest derivative term in the conformal block decomposition of the left diagram has $m_{\rm L}-1$ derivatives of the double-trace conformal block, while for the right diagram it has $m_{\rm R}-1$ derivatives. Explicitly, the coefficients of these highest derivative terms are computed as follows, see (\ref{19jun4})-(\ref{21jun5})
\begin{equation}
\label{28jun1}
\begin{split}
\bar a^{[m_{\rm L}-1]}_{L|\Delta_n}&=-2\pi \lim_{\Delta\to\Delta_n}\left( (\Delta-\Delta_n)^{m_{\rm L}}\bar C_{{\rm L}|\Delta} \right),\\
\bar a^{[m_{\rm R}-1]}_{{\rm R}|\Delta_n}&=-2\pi \lim_{\Delta\to\Delta_n}\left( (\Delta-\Delta_n)^{m_{\rm R}}\bar C_{{\rm R}|\Delta} \right).
\end{split}
\end{equation}
Then, the highest derivative term for the loop diagram has $m_{\rm L}+m_{\rm R}-2$ derivatives and the associated highest derivative coefficient is
\begin{equation}
\label{28jun2}
\begin{split}
\bar a^{[m_{\rm L}+m_{\rm R}-2]}_{{\rm O}|\Delta_n}=-2\pi \lim_{\Delta\to\Delta_n}\left( (\Delta-\Delta_n)^{m_{\rm L}+m_{\rm R}-1}\bar C_{{\rm O}|\Delta} \right).
\end{split}
\end{equation}
Substituting (\ref{7jun7}) into the right hand side of (\ref{28jun2}) and employing (\ref{28jun1}), we obtain
\begin{equation}
\label{28jun3}
\bar a^{[m_{\rm L}+m_{\rm R}-2]}_{{\rm O}|\Delta_n} = \frac{\bar a^{[m_{\rm L}-1]}_{{\rm L}|\Delta_n} \bar a^{[m_{\rm R}-1]}_{{\rm R}|\Delta_n}}{\bar a_{{\rm M}|\Delta_n}},
\end{equation}
where $a_{\rm M}$ denotes the conformal block coefficient for the block of dimension $\Delta_n$ in mean field theory\footnote{These were found in \cite{Dolan:2000ut} in $d=4$ and in \cite{Fitzpatrick:2011dm}  in general dimensions.}. 

Having derived a general formula (\ref{28jun3}), let us now consider particular cases, relevant for typical  one-loop bulk computations.  First, in the case of generic dimensions on the external lines of tree diagrams, each 
$\bar C$ has simple poles at double-trace locations (\ref{27jun6}) and the associated conformal block decompositions do not involve  derivatives of conformal blocks. Then the above analysis implies that the loop diagram also has only simple poles at locations (\ref{27jun6}) and 
(\ref{28jun3}) gives the coefficients of the double-trace conformal blocks of the loop diagram in terms of those for tree-level subdiagrams
\begin{equation}
\label{28jun4}
\bar a^{[0]}_{{\rm O}|\Delta_n} = \frac{\bar a^{[0]}_{{\rm L}|\Delta_n} \bar a^{[0]}_{{\rm L}|\Delta_n}}{\bar a_{{\rm M}|\Delta_n}}.
\end{equation}
For a bubble diagram with no-derivative interaction this formula was found in \cite{Fitzpatrick:2011dm}.

Alternatively, one may consider the case in which  (\ref{25jun4x1}) holds.  In particular, this happens when fields appearing on external lines are identical. Then, second order poles at tree level lead to first derivatives of conformal blocks in the tree-level conformal block decompositions. Moreover, according to the above analysis, the loop diagram will have poles of third order at double-trace locations, resulting in second derivatives of conformal blocks in the conformal block decomposition. The explicit relation between the highest derivative coefficients is given by 
\begin{equation}
\label{28jun5}
\bar a^{[2]}_{{\rm O}|\Delta_n} = \frac{\bar a^{[1]}_{{\rm L}|\Delta_n} \bar a^{[1]}_{{\rm R}|\Delta_n}}{\bar a_{{\rm M}|\Delta_n}}.
\end{equation}
By rewriting this formula in terms of anomalous dimensions, one can show that it is consistent with the expectation from the large-$N$ analysis. This will be further discussed in section \ref{sec:largeN}. Relation (\ref{28jun5}) was also used to compute one-loop diagrams for identical fields, see e. g. \cite{Aharony:2016dwx}.

\subsection{Summary on singular and non-singular contributions}
\label{sec:singsum}
Before finishing this section, let us reiterate once again what our analysis of the double-particle singularity captures and what contributions it misses, now using the language of the conformal block decomposition. 

Firstly, as we explained above, (\ref{28jun3}) captures only the coefficients of the highest derivative terms of conformal blocks of dimensions $\Delta_1+\Delta_2+2n$. In the case (\ref{28jun4}) the highest derivative terms have no derivatives, so (\ref{28jun4}) gives an exact formula for the coefficients of double-trace conformal blocks of dimensions $\Delta_1+\Delta_2+2n$. However, the loop diagram, in addition, contains conformal blocks of dimensions $\Delta_1^{\rm e}+\Delta_2^{\rm e}+2n$ and $\Delta_3^{\rm e}+\Delta_4^{\rm e}+2n$. The associated singularities are present in tree-level subdiagrams due to $B$ factors, see (\ref{25jun2}), and as a consequence of (\ref{6jun12}) also appear for the loop amplitude.
These contributions to the loop amplitude are regular in the sense that these are typical of contact interactions. These regular terms are not captured by (\ref{28jun4}). Alternatively, in the case of identical fields, (\ref{28jun5}) captures the contribution involving second derivatives of double-trace conformal blocks in the conformal block decomposition. The remaining terms -- those with single derivatives and no derivatives of double-trace conformal blocks -- can be regarded as regular as these can be generated by contact interactions. 

Secondly, (\ref{28jun3}) does not capture ''single-trace singularities''. To be more precise, the coefficient functions $A$, see (\ref{25jun2}), for tree-level subdiagrams may contain poles in $\nu$, resulting into single-trace contributions to the conformal block decomposition. This happens, for example, for exchanges in the direct channel. Then, as it is not hard to see from (\ref{6jun12}), these poles carry over to the loop amplitude. Thus, one expects, that the associated single-trace conformal blocks are also present in the conformal block decomposition of the loop amplitude. Similarly, these single-trace contributions to the loop amplitude are not captured by our analysis. At the same time, it is worth stressing, that the presence/absence of single-trace blocks for tree-level subdiagrams does not affect the result (\ref{28jun3}) for double-trace conformal blocks.

The situation we described here is identical to that in  flat space, in which the double-particle cut diagram captures the associated discontinuity of the amplitude, but does not say anything about the regular part of the amplitude and singularities associated with other cut diagrams.

\section{Large-$N$ expansion for the dual theory}
\label{sec:largeN}

In the previous section we demonstrated how the double-cut singularity  of a one-loop bulk diagram is related to tree-level diagrams obtained from the original loop diagram by cutting two propagators. In this section we will review the CFT dual counterpart of this analysis. Our goal is to demonstrate that relations such as (\ref{28jun4}) and (\ref{28jun5}) 
are consistent with the standard large-$N$ considerations.
As on the bulk side, we will have two different situations: the one in which the degeneracy condition (\ref{25jun4x1}) is satisfied
 and the other one in which it is not. In these two cases the analysis is somewhat different. Let us start from a non-degenerate case. 

\subsection{Non-degenerate case}

In this section we will be interested in a one-loop diagram in AdS space, which has the fields dual to operators ${\cal O}_{1}^{\rm e}$, ${\cal O}_{2}^{\rm e}$, ${\cal O}_{3}^{\rm e}$ and ${\cal O}_{4}^{\rm e}$ on external lines and a non-trivial two-particle cut through lines propagating fields dual to operators ${\cal O}_1$ and ${\cal O}_2$. Our goal is to relate the CFT data of the singular part of the loop diagram associated with a given cut to the CFT data of the tree diagrams, resulting from the cut of the loop diagram, employing CFT considerations. Moreover, we will assume that dimensions of the fields are such that degeneracy condition (\ref{25jun4x1}) never takes place.

We begin by analyzing the CFT dual of tree-level amplitudes appearing after cutting  a loop diagram. These correspond to 
correlators
\begin{equation}
\label{1jul1}
\langle {\cal O}_1^{\rm e} {\cal O}_2^{\rm e} {\cal O}_1 {\cal O}_2 \rangle, 
\qquad
\langle {\cal O}_3^{\rm e} {\cal O}_4^{\rm e} {\cal O}_1 {\cal O}_2 \rangle,
\end{equation}
where ${\cal O}_i^{\rm e}$ and ${\cal O}_i$ are the single-trace operators dual to the bulk fields appearing on external lines and running in the loop of the bulk amplitude respectively. Expanding the first correlator at large $N$, we find
\begin{equation}
\label{1jul2}
\langle {\cal O}_1^{\rm e} {\cal O}_2^{\rm e} {\cal O}_1 {\cal O}_2 \rangle
=\langle {\cal O}_1^{\rm e} {\cal O}_2^{\rm e} {\cal O}_1 {\cal O}_2 \rangle^{(0)}+
\frac{1}{N^2}\langle {\cal O}_1^{\rm e} {\cal O}_2^{\rm e} {\cal O}_1 {\cal O}_2 \rangle^{(1)}+\dots.
\end{equation}
Here the leading term in the expansion vanishes, as it corresponds to the disconnected diagram, which is absent in the case in which all operators are different. By AdS/CFT correspondence, the bulk gravitational constant $G_N$ equals $1/N^2$\footnote{Formula $G_N=1/N^2$ at large $N$ refers to particular examples of the holographic correspondence, such as the classical one  \cite{Maldacena:1997re,Witten:1998qj,Gubser:1998bc}. Instead, in the present paper we rather discuss general bulk theories, which may have many independent coupling constants. In this case we require that $g_3\propto 1/N$, $g_4\propto 1/N^2$ for cubic and quartic couplings in the large-$N$ limit and similarly for higher order interactions. This scaling guarantees the appropriate identification of the bulk loop expansion and the $1/N$ expansion on the boundary.} at large $N$, so the second term on the right hand side of (\ref{1jul2}) corresponds to the tree-level contribution we are interested in.

The OPE of ${\cal O}_1^{\rm e}$ and ${\cal O}_2^{\rm e}$ has the schematic form
\begin{equation}
\label{1jul3}
\begin{split}
{\cal O}_1^{\rm e}\times {\cal O}_2^{\rm e}&=
\left(\frac{1}{N} +\dots \right){\cal O}_i+ \left(1+\frac{1}{N^2} +\dots \right)[{\cal O}_1^{\rm e}{\cal O}_2^{\rm e}]_{n,l}\\
&\qquad\qquad \qquad\qquad+
\left(\frac{1}{N^2} +\dots \right)[{\cal O}_1{\cal O}_2]_{n,l}+\dots.
\end{split}
\end{equation}
Here ${\cal O}_i$ refers to all single-trace operators that can appear in a given OPE, while  $[{\cal O}_i{\cal O}_j]_{n,l}$ and
$[{\cal O}^{\rm e}_i{\cal O}^{\rm e}_j]_{n,l}$
are the double-trace operators of the schematic form
\begin{equation}
\label{1jul4}
[{\cal O}_i{\cal O}_j]_{n,l} = {\cal O}_i \Box^n \partial_{\mu_1}\dots\partial_{\mu_l}{\cal O}_j+\dots
\end{equation}
and similarly for $[{\cal O}^{\rm e}_i{\cal O}^{\rm e}_j]_{n,l}$.
At large $N$ they have the following dimensions
\begin{equation}
\label{1jul5x1}
\Delta_{ij|n,l} = \Delta_{ij|n,l}^{(0)}+\frac{1}{N^2} \gamma_{ij|n,l}^{(1)}+\dots, \qquad 
\Delta^{\rm ee}_{ij|n,l} = \Delta_{ij|n,l}^{{\rm ee}|(0)}+\frac{1}{N^2} \gamma_{ij|n,l}^{{\rm ee}|(1)}+\dots,
\end{equation}
where
\begin{equation}
\label{1jul5}
\Delta_{ij|n,l}^{(0)}=\Delta_i+\Delta_j+2n+l, \qquad \Delta_{ij|n,l}^{{\rm ee}|(0)}=\Delta^{\rm e}_i+\Delta^{\rm e}_j+2n+l.
\end{equation}
and $\gamma$ are the anomalous dimensions. 

The OPE coefficients implicitly appearing in (\ref{1jul3}) also admit the $1/N$ expansion, for which we introduce the following notations
\begin{equation}
\label{1jul6}
\begin{split}
c_{{\cal O}_1^{\rm e} {\cal O}_2^{\rm e} {\cal O}_i} &= \frac{1}{N}c^{(1)}_{{\cal O}_1^{\rm e} {\cal O}_2^{\rm e} {\cal O}_i}+\dots ,\\
c_{{\cal O}_1^{\rm e} {\cal O}_2^{\rm e} [{\cal O}_1^{\rm e}{\cal O}_2^{\rm e}]_{n,l}} &=
c^{(0)}_{{\cal O}_1^{\rm e} {\cal O}_2^{\rm e} [{\cal O}_1^{\rm e}{\cal O}_2^{\rm e}]_{n,l}} +\frac{1}{N^2}
c^{(1)}_{{\cal O}_1^{\rm e} {\cal O}_2^{\rm e} [{\cal O}_1^{\rm e}{\cal O}_2^{\rm e}]_{n,l}} +\dots ,\\
c_{{\cal O}_1^{\rm e} {\cal O}_2^{\rm e} [{\cal O}_1{\cal O}_2]_{n,l}} &=\frac{1}{N^2}
c^{(1)}_{{\cal O}_1^{\rm e} {\cal O}_2^{\rm e} [{\cal O}_1{\cal O}_2]_{n,l}} +\dots.
\end{split}
\end{equation}
Note the appearance of $c^{(0)}$, which are the OPE coefficients of  mean field theory. 
The OPE of ${\cal O}_1$ and ${\cal O}_2$  is analogous.

In these terms, the conformal block decomposition of $\langle {\cal O}_1^{\rm e} {\cal O}_2^{\rm e} {\cal O}_1 {\cal O}_2 \rangle$
acquires the form\footnote{Hopefully, the notation in which we write the operator itself instead of its dimension as an index of a conformal block will not lead to any confusions.}
\begin{equation}
\label{1jul7}
\begin{split}
\langle {\cal O}_1^{\rm e} {\cal O}_2^{\rm e} {\cal O}_1 {\cal O}_2 \rangle^{(1)}&=
\sum_i  c^{(1)}_{{\cal O}_1^{\rm e} {\cal O}_2^{\rm e} {\cal O}_i} c^{(1)}_{{\cal O}_1 {\cal O}_2 {\cal O}_i}G^{\Delta^{\rm e}_1\Delta^{\rm e}_2\Delta_1\Delta_2}_{\Delta_i}\\
&+\sum_{n,l} c^{(0)}_{{\cal O}_1^{\rm e} {\cal O}_2^{\rm e} [{\cal O}_1^{\rm e}{\cal O}_2^{\rm e}]_{n,l}}c^{(1)}_{{\cal O}_1 {\cal O}_2 [{\cal O}_1^{\rm e}{\cal O}_2^{\rm e}]_{n,l}}G^{\Delta^{\rm e}_1\Delta^{\rm e}_2\Delta_1\Delta_2}_{[{\cal O}_1^{\rm e}{\cal O}_2^{\rm e}]_{n,l}}\\
&+\sum_{n,l} c^{(1)}_{{\cal O}_1^{\rm e} {\cal O}_2^{\rm e} [{\cal O}_1{\cal O}_2]_{n,l}}c^{(0)}_{{\cal O}_1 {\cal O}_2 [{\cal O}_1{\cal O}_2]_{n,l}}G^{\Delta^{\rm e}_1\Delta^{\rm e}_2\Delta_1\Delta_2}_{[{\cal O}_1{\cal O}_2]_{n,l}}.
\end{split}
\end{equation}
It is not hard to see that this conformal block decomposition is consistent with the one we encountered for the bulk tree-level diagrams in section \ref{sec:4pt}. Indeed, the first line in (\ref{1jul7}) contains single-trace conformal blocks generated by the $A$ factor in (\ref{25jun2}), while the other two lines are the double-trace contributions generated by the kinematic $B$ factor. Analogous relations hold for the second 
correlator in (\ref{1jul1}) at the leading order --- one just needs to replace ${\cal O}^{\rm e}_1$ and ${\cal O}^{\rm e}_2$ with 
${\cal O}^{\rm e}_3$ and ${\cal O}^{\rm e}_4$.

Now, let us move to the correlator $\langle {\cal O}_1^{\rm e} {\cal O}_2^{\rm e} {\cal O}^{\rm e}_3 {\cal O}^{\rm e}_4 \rangle$ and focus on its $O(1/N^4)$ part, that corresponds to the one-loop level. To derive its conformal block decomposition, we need to take into account contributions from all operators that appear simultaneously in the  OPE's ${\cal O}_1^{\rm e} \times{\cal O}_2^{\rm e}$ and ${\cal O}^{\rm e}_3 \times {\cal O}^{\rm e}_4 $ with the OPE coefficients and anomalous dimensions, that are relevant at this order. 

First, we consider $[{\cal O}_1{\cal O}_2]_{n,l}$. They appear in both OPE's with the leading OPE coefficient of order $O(1/N^2)$, see (\ref{1jul6}). This means that we should have the following contribution
\begin{equation}
\label{1jul8}
\langle {\cal O}_1^{\rm e} {\cal O}_2^{\rm e} {\cal O}^{\rm e}_3 {\cal O}^{\rm e}_4 \rangle^{(2)} \supset
\sum_{n,l} c^{(1)}_{{\cal O}_1^{\rm e} {\cal O}_2^{\rm e} [{\cal O}_1{\cal O}_2]_{n,l}}c^{(1)}_{{\cal O}_3^{\rm e} {\cal O}_4^{\rm e} [{\cal O}_1{\cal O}_2]_{n,l}}G^{\Delta^{\rm e}_1\Delta^{\rm e}_2\Delta_3^{\rm e}\Delta_4^{\rm e}}_{[{\cal O}_1{\cal O}_2]_{n,l}}.
\end{equation}
This is precisely the contribution we are after. It is not hard to see that the conformal block coefficients appearing in (\ref{1jul8}) are related to the conformal block coefficients in the last line of  (\ref{1jul7}) and the analogous ones for $\langle {\cal O}_3^{\rm e} {\cal O}_4^{\rm e} {\cal O}_1 {\cal O}_2 \rangle^{(1)}$ as found in (\ref{28jun4})\footnote{On the bulk side we only discussed contributions of scalar double-trace operators, $l=0$, and the spin label was omitted.}.

Other set of operators that should be taken into account is $[{\cal O}^{\rm e}_1{\cal O}^{\rm e}_2]_{n,l}$. These may give two types of contributions. The first type  is
\begin{equation}
\label{1jul9}
\langle {\cal O}_1^{\rm e} {\cal O}_2^{\rm e} {\cal O}^{\rm e}_3 {\cal O}^{\rm e}_4 \rangle^{(2)} \supset
\sum_{n,l} c^{(0)}_{{\cal O}_1^{\rm e} {\cal O}_2^{\rm e} [{\cal O}^{\rm e}_1{\cal O}^{\rm e}_2]_{n,l}}c^{(2)}_{{\cal O}_3^{\rm e} {\cal O}_4^{\rm e} [{\cal O}^{\rm e}_1{\cal O}^{\rm e}_2]_{n,l}}G^{\Delta^{\rm e}_1\Delta^{\rm e}_2\Delta_3^{\rm e}\Delta_4^{\rm e}}_{[{\cal O}^{\rm e}_1{\cal O}^{\rm e}_2]_{n,l}}.
\end{equation}
These terms are, however, regular and we are not interested in them here. Another type of contributions is of the form
\begin{equation}
\label{1jul10}
\langle {\cal O}_1^{\rm e} {\cal O}_2^{\rm e} {\cal O}^{\rm e}_3 {\cal O}^{\rm e}_4 \rangle^{(2)} \supset
\sum_{n,l} c^{(0)}_{{\cal O}_1^{\rm e} {\cal O}_2^{\rm e} [{\cal O}^{\rm e}_1{\cal O}^{\rm e}_2]_{n,l}}c^{(1)}_{{\cal O}_3^{\rm e} {\cal O}_4^{\rm e} [{\cal O}^{\rm e}_1{\cal O}^{\rm e}_2]_{n,l}}
\gamma_{12|n,l}^{{\rm ee}|(1)}
\frac{\partial}{\partial \Delta_{12|n,l}^{{\rm ee}}}G^{\Delta^{\rm e}_1\Delta^{\rm e}_2\Delta_3^{\rm e}\Delta_4^{\rm e}}_{[{\cal O}^{\rm e}_1{\cal O}^{\rm e}_2]_{n,l}}.
\end{equation}
It is linear in $c^{(1)}_{{\cal O}_3^{\rm e} {\cal O}_4^{\rm e} [{\cal O}^{\rm e}_1{\cal O}^{\rm e}_2]_{n,l}}$, which is the CFT data of the tree-level correlator $\langle {\cal O}_1^{\rm e} {\cal O}_2^{\rm e} {\cal O}^{\rm e}_3 {\cal O}^{\rm e}_4 \rangle^{(1)}$. It is also linear in $\gamma_{12|n,l}^{{\rm ee}|(1)}$, which is the CFT data of the tree-level correlator $\langle {\cal O}_1^{\rm e} {\cal O}_2^{\rm e} {\cal O}^{\rm e}_1 {\cal O}^{\rm e}_2 \rangle^{(1)}$. This implies, that (\ref{1jul10}) is the contribution associated with a double-particle cut through lines, that propagate fields dual to ${\cal O}_1^{\rm e}$ and ${\cal O}_2^{\rm e}$. This diagram may be non-vanishing, but it is not the diagram we set to compute here.

Finally, let us consider contributions from single-trace operators ${\cal O}_i$. These can also be of two types
\begin{equation}
\label{1jul12}
\begin{split}
\langle {\cal O}_1^{\rm e} {\cal O}_2^{\rm e} {\cal O}^{\rm e}_3 {\cal O}^{\rm e}_4 \rangle^{(2)} \supset 
\sum_i  \left( c^{(2)}_{{\cal O}_1^{\rm e} {\cal O}_2^{\rm e} {\cal O}_i} c^{(1)}_{{\cal O}_3^{\rm e} {\cal O}_4^{\rm e} {\cal O}_i}
+
 c^{(1)}_{{\cal O}_1^{\rm e} {\cal O}_2^{\rm e} {\cal O}_i} c^{(2)}_{{\cal O}_3^{\rm e} {\cal O}_4^{\rm e} {\cal O}_i}
\right)
G^{\Delta^{\rm e}_1\Delta^{\rm e}_2\Delta_3^{\rm e}\Delta_4^{\rm e}}_{\Delta_i}\\
+
\sum_i  c^{(1)}_{{\cal O}_1^{\rm e} {\cal O}_2^{\rm e} {\cal O}_i} c^{(1)}_{{\cal O}_3^{\rm e} {\cal O}_4^{\rm e} {\cal O}_i}
\gamma_{i}^{(1)}\frac{\partial}{\partial \Delta_i}
G^{\Delta^{\rm e}_1\Delta^{\rm e}_2\Delta_3^{\rm e}\Delta_4^{\rm e}}_{\Delta_i}.
\end{split}
\end{equation}
From the bulk perspective these contributions correspond to single-particle cuts. Indeed, the first line involves $c^{(2)}$ for single-trace operators, which corresponds to one-loop corrections to the bulk cubic vertex. The second line involves anomalous dimensions $\gamma^{(1)}$ for the single-trace operators, which via holography maps to the mass shift of the bulk propagator. Such contributions cannot be derived from unitarity and analyticity in flat space and also they are not captured by the analysis of AdS double-cut diagrams in the previous section. On the CFT side these cannot be reconstructed via large-$N$ expansion neither, but rather should be taken as an input data.

\subsection{Degenerate case}
\label{sec:deg}

Here we review the CFT counterpart of the setting for which degeneracy (\ref{25jun4x1}) occurs. For simplicity, we consider 
the case, in which all operators -- appearing on the external lines and on the cut lines -- are identical.

The OPE of ${\cal O}$ with itself is schematically of the form
\begin{equation}
\label{25jun6}
{\cal O}\times {\cal O} = {\bf 1}+ \left(\frac{1}{N} +\dots \right){\cal O} + \left(1+\frac{1}{N^2} +\dots \right)[{\cal O}{\cal O}]_{n,l}+\dots.
\end{equation}
Here ${\bf 1}$ denotes the identity operator and $[{\cal O}{\cal O}]_{n,l}$ are the double-trace operators as defined in (\ref{1jul4}). As in the previous section, we expand the CFT data in $1/N$, which also specifies the four-point correlator order by order. The role of single-trace operators in this discussion is the same as in the previous section, so below we focus on the double-trace contributions only.

  At order $O(N^0)$ the double-trace operators $[{\cal O}{\cal O}]_{n,l}$ have the mean field theory dimensions
\begin{equation}
\label{25jun8}
\Delta_{n,l}^{(0)}=2\Delta+2n+l,
\end{equation}
which then acquire corrections 
\begin{equation}
\label{25jun9}
\Delta_{n,l}=\Delta^{(0)}_{n,l}+\frac{1}{N^2}\gamma^{(1)}_{n,l}+\frac{1}{N^4}\gamma_{n,l}^{(2)}+\dots.
\end{equation}
Similarly, the OPE coefficients for $[{\cal O}{\cal O}]_{n,l}$ in (\ref{25jun6}) admit the $1/N$ expansion
\begin{equation}
\label{25jun10}
c_{{\cal O}{\cal O}[{\cal O}{\cal O}]_{n,l}} = c^{(0)}_{{\cal O}{\cal O}[{\cal O}{\cal O}]_{n,l}}+\frac{1}{N^2}
c^{(1)}_{{\cal O}{\cal O}[{\cal O}{\cal O}]_{n,l}}+\frac{1}{N^4}c^{(2)}_{{\cal O}{\cal O}[{\cal O}{\cal O}]_{n,l}}+\dots
\end{equation}
and $c^{(0)}$ refers to the OPE coefficients in mean field theory.

Accordingly, the four point correlator can be expanded as
\begin{equation}
\label{26jun1}
\langle {\cal O}{\cal O}{\cal O}{\cal O} \rangle = 
\langle {\cal O}{\cal O}{\cal O}{\cal O} \rangle^{(0)}+\frac{1}{N^2}\langle {\cal O}{\cal O}{\cal O}{\cal O} \rangle^{(1)}+
\frac{1}{N^4}\langle {\cal O}{\cal O}{\cal O}{\cal O} \rangle^{(2)}+\dots.
\end{equation}
Here the leading term is just the disconnected correlator. 
The subleading term for the four-point correlator admits the conformal block decomposition
\begin{equation}
\label{26jun2}
\langle {\cal O}{\cal O}{\cal O}{\cal O} \rangle^{(1)}=
2 c^{(0)}_{{\cal O}{\cal O}[{\cal O}{\cal O}]_{n,l}} c^{(1)}_{{\cal O}{\cal O}[{\cal O}{\cal O}]_{n,l}} G^{\Delta}_{\Delta_{n,l}}
+\left(c^{(0)}_{{\cal O}{\cal O}[{\cal O}{\cal O}]_{n,l}} \right)^2\gamma^{(1)}_{n,l}
\frac{\partial}{\partial \Delta_{n,l}}G^{\Delta}_{\Delta_{n,l}}.
\end{equation}
Proceeding to  order $1/N^4$ we find the following terms
\begin{equation}
\label{26jun4}
\begin{split}
\langle {\cal O}{\cal O}{\cal O}{\cal O} \rangle^{(2)}=&
\left(2 c^{(0)}_{{\cal O}{\cal O}[{\cal O}{\cal O}]_{n,l}} c^{(2)}_{{\cal O}{\cal O}[{\cal O}{\cal O}]_{n,l}} +
\left(c^{(1)}_{{\cal O}{\cal O}[{\cal O}{\cal O}]_{n,l}} \right)^2
\right)
G^{\Delta}_{\Delta_{n,l}}\\
+&\left(\left(c^{(0)}_{{\cal O}{\cal O}[{\cal O}{\cal O}]_{n,l}} \right)^2\gamma^{(2)}_{n,l}
+2 c^{(0)}_{{\cal O}{\cal O}[{\cal O}{\cal O}]_{n,l}} c^{(1)}_{{\cal O}{\cal O}[{\cal O}{\cal O}]_{n,l}} 
\gamma^{(1)}_{n,l}
\right)\frac{\partial}{\partial \Delta_{n,l}}G^{\Delta}_{\Delta_{n,l}}
\\
+&\frac{1}{2}\left(c^{(0)}_{{\cal O}{\cal O}[{\cal O}{\cal O}]_{n,l}}\gamma^{(1)}_{n,l} \right)^2
\frac{\partial^2}{\partial \Delta^2_{n,l}}G^{\Delta}_{\Delta_{n,l}}.
\end{split}
\end{equation}
Here the term in the last line is the only singular contribution. Moreover, it is completely fixed by order $1/N^2$ CFT data. 
It is not hard to see that the relation between the coefficients of second  derivatives of double-trace conformal blocks in (\ref{26jun4}) and the coefficients of first derivatives in (\ref{26jun2}) is given by (\ref{28jun5}), that we found from bulk considerations.

\section{Generalizations and consequences}
\label{sec:gen}

In this section we will comment on a number of straightforward extensions and corollaries of the previous discussion.

\subsection{Higher-loop and higher-point functions}

The arguments presented above admit a straightforward extension to higher-point and higher-loop amplitudes. Let us illustrate this with a simple example. 

Suppose we have a four-point two-loop amplitude which admits a three-particle cut in a given channel. Using the split representation for the propagators involved in the cut, we rewrite a given amplitude as an integrated product of two tree-level five-point functions. By appropriately choosing the channel for the tree-level amplitudes we find, schematically,
\begin{equation}
\label{4jul1}
\begin{split}
A_{\rm O}(Q_i) &= \int dP_{{\rm L}_1}dP_{{\rm L}_2}dP_{{\rm R}_1}dP_{{\rm R}_2}dP_1dP_2dP_3
d\nu_{{\rm L}_1}d\nu_{{\rm L}_2}d\nu_{{\rm R}_1}d\nu_{{\rm R}_2}d\nu_1d\nu_2d\nu_3 (\dots)\\
&
[{\cal O}_{\Delta^{\rm e}_1}(Q_1) {\cal O}_{\Delta^{\rm e}_2}(Q_2) {\cal O}_{\Delta_{{\rm L}_1}}(P_{{\rm L}_1})]
[ {\cal O}_{\tilde\Delta_{{\rm L}_1}}(P_{{\rm L}_1}) {\cal O}_{h+i\nu_1}(P_1) {\cal O}_{\Delta_{{\rm L}_2}}(P_{{\rm L}_2}) ]\\
&\quad\quad
[ {\cal O}_{\tilde\Delta_{{\rm L}_2}}(P_{{\rm L}_2}) {\cal O}_{h+i\nu_2}(P_2) {\cal O}_{h+i\nu_{3}}(P_{3}) ]
[ {\cal O}_{h-i\nu_2}(P_2) {\cal O}_{h-i\nu_{3}}(P_{3}) {\cal O}_{\Delta_{{\rm R}_2}}(P_{{\rm R}_2}) ]\\
&\qquad\qquad
[  {\cal O}_{\tilde\Delta_{{\rm R}_2}}(P_{{\rm R}_2})
{\cal O}_{h-i\nu_1}(P_1) {\cal O}_{\Delta_{{\rm R}_1}}(P_{{\rm R}_1}) ]
[   {\cal O}_{\tilde\Delta_{{\rm R}_1}}(P_{{\rm R}_1})
{\cal O}_{\Delta_3^{\rm e}}(Q_3) {\cal O}_{\Delta_4^{\rm e}}(Q_4) ].
\end{split}
\end{equation}
Here $P_1$, $P_2$ and $P_3$ are additional boundary points introduced by the split representation of the propagators to be cut and $\nu_1$, $\nu_2$ and $\nu_3$ are the associated spectral parameters. Analogously,  $P_{{\rm L}_1}$,  $P_{{\rm L}_2}$, $P_{{\rm R}_1}$ and $P_{{\rm R}_2}$ are the intermediate points of conformal partial waves for the left and the right tree amplitudes and $\nu_{{\rm L}_1}$,  $\nu_{{\rm L}_2}$, $\nu_{{\rm R}_1}$ and $\nu_{{\rm R}_2}$ are the respective spectral parameters. Various coefficient functions in (\ref{4jul1})  we leave implicit.

We can analyze (\ref{4jul1}) using iterative applications of the procedure from the previous section. To be more precise, we start by evaluating $P_2$ and $P_3$ integrals employing the bubble integral formula. Then delta-functions that it produces can be used to remove $\nu_{{\rm R}_2}$ and $P_{{\rm R}_2}$ integrals. After that we integrate out $P_1$ and $P_{{\rm L}_2}$ using the bubble integral formula again, which, eventually, leads to the conformal partial wave expansion for the original four-point diagram. 

Similarly, one can use the arguments of the previous sections iteratively to study the analytic structure of the resulting spectral integrals. Namely, we first consider the integral over $\nu_2$ and $\nu_3$ and find singularities in $\nu_{{\rm L}_2}$ that it produces. Next, we use the same methods to find singularities in $\nu_{{\rm L}_1}$ after $\nu_{{\rm L}_2}$ and $\nu_1$ are integrated out. Eventually, we find the triple-cut -- or, equivalently, triple-trace -- singularities of the conformal partial wave expansion of the initial two-loop diagram.

In summary, to deal with amplitudes that admit non-trivial cuts of multiple propagators, one should expand each subdiagram in partial waves and then proceed iteratively, at each step applying the bubble formula to a bubble formed by a pair of propagators. Each iteration, effectively, reduces the number of propagators in the diagram by one: instead of a pair of propagators with dimensions $\Delta_1$ and $\Delta_2$ one obtains a single propagator with  singularities at double-trace locations $\Delta_1+\Delta_2+2n$. The iterative procedure stops when there is only one propagator left. Its singularities correspond to the multiple-trace operators formed from the operators associated with the cut lines of the original diagram. This procedure is analogous to the approach used in \cite{Fitzpatrick:2011hu} to compute higher loop bubble diagrams. However, unlike the approach of \cite{Fitzpatrick:2011hu}, our procedure is applicable to all types of diagrams.

These results can be compared with the large-$N$ expansion in the CFT. The leading  contribution to the five-point correlator is of order $O(1/N^3)$. Expanding the left correlator into conformal blocks in the same channel as the left amplitude in (\ref{4jul1}), we will find the following terms
\begin{equation}
\label{4jul2}
\begin{split}
\langle {\cal O}_1^{\rm e}{\cal O}_2^{\rm e} {\cal O}_1 {\cal O}_2 {\cal O}_3\rangle^{(1)} \supset
c^{(1)}_{
{\cal O}_1^{\rm e} {\cal O}_2^{\rm e} [{\cal O}_1{\cal O}_2{\cal O}_3]
 }
 c^{(0)}_{
 [{\cal O}_1{\cal O}_2{\cal O}_3]{\cal O}_1 [{\cal O}_2{\cal O}_3]
 }
 c^{(0)}_{[{\cal O}_2{\cal O}_3]{\cal O}_2{\cal O}_3}
 G^{\Delta_1^{\rm e}\Delta_2^{\rm e}\Delta_1\Delta_2\Delta_3}_{ [{\cal O}_1{\cal O}_2{\cal O}_3],  [{\cal O}_2{\cal O}_3]}
\end{split}
\end{equation}
and similarly for the right amplitude. Then, applying bulk formula (\ref{28jun4}) two times we find that the two-loop amplitude has singular terms associated with a three-particle cut of the form
\begin{equation}
\label{4jul3}
\langle {\cal O}_1^{\rm e}{\cal O}_2^{\rm e} {\cal O}_3^{\rm e}{\cal O}_4^{\rm e} \rangle \supset 
c^{(1)}_{
{\cal O}_1^{\rm e} {\cal O}_2^{\rm e} [{\cal O}_1{\cal O}_2{\cal O}_3]
 }
 c^{(1)}_{
{\cal O}_3^{\rm e} {\cal O}_4^{\rm e} [{\cal O}_1{\cal O}_2{\cal O}_3]
 }
 G^{\Delta_1^{\rm e} \Delta_2^{\rm e} \Delta_3^{\rm e} \Delta_4^{\rm e}}_{[{\cal O}_1{\cal O}_2{\cal O}_3]},
\end{equation}
which is consistent with the expectation form the CFT analysis. 

Clearly, this argument can be generalized to any number of cut propagators and any number of external lines for the diagrams involved. The same procedure can also be applied for the case in which the diagrams resulting from a cut involve loops themselves. More thorough and systematic analysis of these extensions we leave for future research.

\subsection{AdS Cutkosky rules}

The effect of the contour deformation used in section \ref{sec:sing} to extract the singular part of the amplitude, eventually, amounts to the replacement of propagators with cut propagators
\begin{equation}
\label{5jul1}
\Pi_{\Delta}(X_1,X_2)\quad \to\quad  \frac{2\pi}{\Delta-h}\Omega_{i(\Delta-h)} (X_1,X_2)= \Pi_\Delta(X_1,X_2)-\Pi_{d-\Delta}(X_1,X_2).
\end{equation}
This can be regarded as the AdS version of flat space Cutkosky rules \cite{Cutkosky:1960sp,smatrix}\footnote{The analogy between $\Pi_\Delta-\Pi_{d-\Delta}$ and flat space cut propagators is rather obvious: indeed, they both satisfy free equations of motion identically. Moreover, the fact that substitution $\Pi_\Delta\to \Pi_\Delta-\Pi_{d-\Delta}$ into the exchange diagram produces only a single-trace conformal block with its shadow partner is well-known. In (\ref{5jul1}) we state that this idea naturally extends to loop amplitudes.}. Let us remind the reader again, that in contrast to the usual Cutkosky rules, replacement (\ref{5jul1}) not only gives the singular part of the amplitude, but also results in additional shadow singularities, which were not present in the initial amplitude. It would be interesting to understand how these shadow contributions can be removed in future. Presumably, this should require analytic continuation of the amplitude to the Lorentzian signature. Alternatively, one can project out the unnecessary contributions using monodromy transformations, see \cite{SimmonsDuffin:2012uy}.

It is also instructive to reformulate the prescription (\ref{5jul1}) in terms of the CFT correlators. By taking care of all the necessary normalization factors, for the double-particle cut we find that 
\begin{equation}
\label{5jul2}
\begin{split}
&\langle{\cal O}_1^{\rm e}(Q_1) {\cal O}_2^{\rm e}(Q_2) {\cal O}_3^{\rm e}(Q_3) {\cal O}_4^{\rm e}(Q_4) \rangle_{\rm O}
 \supset
\langle{\cal O}_1^{\rm e}(Q_1) {\cal O}_2^{\rm e}(Q_2) {\cal O}_1(P_1) {\cal O}_2(P_2) \rangle_{\rm L}\\
&\qquad\qquad \frac{1}{{\cal N}_{\Delta_1}}\frac{1}{{\cal N}_{\Delta_2}} [ \tilde{\cal O}_1(P_1) \tilde{\cal O}_1 (P'_1)]
[ \tilde{\cal O}_2(P_2) \tilde{\cal O}_2(P'_2) ]
\langle{\cal O}_3^{\rm e}(Q_3) {\cal O}_4^{\rm e}(Q_4) {\cal O}_1(P'_1) {\cal O}_2(P'_2) \rangle_{\rm R},
\end{split}
\end{equation}
where
\begin{equation}
\label{5jul3}
{\cal N}_\Delta \equiv \frac{\pi^d \Gamma(\Delta-h)\Gamma(h-\Delta)}{\Gamma(\Delta)\Gamma(d-\Delta)}.
\end{equation}
The operation of insertion of the operator
\begin{equation}
\label{5jul4}
|  {\cal O}(P_1)\rangle \frac{1}{{\cal N}_\Delta} [ \tilde{\cal O}_1(P_1) \tilde{\cal O}_1 (P'_1)] \langle {\cal O}(P_2) |
\end{equation}
into a correlator is known to carry out its projection onto a contribution associated with the operator ${\cal O}$ and its shadow \cite{SimmonsDuffin:2012uy}. In particular, when inserted into a four-point correlator, it gives a contribution associated with the conformal block in which  the operator ${\cal O}$ is exchanged plus its shadow partner. Similarly, multiple insertions of the projector (\ref{5jul4}) as they appear in (\ref{5jul2}) can be understood as a projection of the correlator onto the space of multi-particle states for the associated set of operators. This interpretation parallels the one of the Cutkosky rules, in which a particular singularity of the $S$-matrix is expressed in terms of an integral over the on-shell phase space of particles associated with the cut propagators. A closely related discussion in a somewhat different form can be found in \cite{Fitzpatrick:2011dm}.

\subsection{Spinning fields}
\label{sec:sf}

So far our analysis was focused on scalar fields only. The procedure we employed, however, can be straightforwardly generalized to include fields with spin. The main technical difficulty related to such a generalization is due to the presence of multiple tensor structures for three-point correlators of operators of general spin and due to the necessity to compute all possible bubble integrals involving these tensor structures.

The simplest extension to consider along these lines is to take into account spinning conformal blocks with scalar operators on external lines. These contributions are relevant even in theories of scalar fields in the bulk, if, for example, a quartic vertex contains derivatives or cubic couplings are non-vanishing. To be able to compute the singular part of a one-loop amplitude using the procedure from sections \ref{sec:loopfromtree} and \ref{sec:sing}, one needs to deal with bubble integrals that involve two spinning operators on external lines and two scalar operators at the points being integrated out. Such bubble integrals are known \cite{Dobrev:1976vr,Karateev:2018oml}. It is straightforward to check that this computation eventually leads to the result of the form (\ref{28jun3}), in which we just need to replace conformal block coefficients for scalar operators with spinning ones.

Instead of giving this computation explicitly, we will present a shortcut method. 
It is clear, that if the loop computation is done directly, we will find a relation of the form
\begin{equation}
\label{5jul5}
a^{[0]}_{{\rm O}|\Delta_{n,l},l} = \alpha_{n,l}{a^{[0]}_{{\rm L}|\Delta_{n,l},l} a^{[0]}_{{\rm R}|\Delta_{n,l},l}},
\end{equation}
in which it only remains to find the multiplicative factor $\alpha_{n,l}$. To do that, we will use that the singular part of the loop diagram can be computed by the sewing procedure (\ref{5jul2}). This formula is valid for any correlators used in place of tree-level amplitudes and we will take them to be the disconnected correlators. Then, due to the standard identities with the two-point correlators, the left hand side of (\ref{5jul2}) is also a disconnected correlator. Thus, plugging for all $a$'s in (\ref{5jul5}) the conformal block coefficients of mean field theory, we find that 
\begin{equation}
\label{5jul6}
\alpha_{n,l} = \left(a^{[0]}_{{\rm M}|\Delta_{n,l},l}\right)^{-1},
\end{equation}
which was to be demonstrated. Generalization to the case in which anomalous dimensions are present is straightforward. The relation between the bubble integral and the mean field theory conformal block coefficients that we encountered here in a similar context appeared previously in \cite{Karateev:2018oml}.

For more general spin configurations the same argument can be used to bypass the computation of the bubble integral explicitly. It is worth stressing, however, that only particular types of tensor structures appear in three-point correlators of mean field theory, so only some of the cut diagrams in the bulk can be computed with this trick. If the tree amplitudes resulting from a cut of a loop diagram involve different tensor structures in their conformal block decompositions, then the associated bubble diagrams have to be evaluated explicitly. It would be interesting to see what is the analogue of (\ref{28jun3}) in this case and how this procedure can be reconciled with the large-$N$ expansion for the CFT dual theory. 

\subsection{Reconstruction of the complete amplitude}

Here we briefly mention the issue of reconstruction of the complete amplitude from its singularities. In flat space, once the high-energy behavior is known, this reconstruction can be carried out using simple arguments from complex analysis. Analogous approaches were recently developed in the CFT  \cite{Alday:2016njk,Caron-Huot:2017vep,Alday:2017vkk,Simmons-Duffin:2017nub}. Similarly to the flat space approach, in the CFT's one can derive bounds on the correlators in the Regge limit both at finite $N$ \cite{Caron-Huot:2017vep} and in the large-$N$ limit \cite{Maldacena:2015waa} and then, using analyticity, reconstruct the complete correlator from singular terms up to a finite number of lower-spin contributions compatible with a given Regge behavior.

This analysis is particularly simple when applied to the correlators or AdS amplitudes in the form of the conformal partial wave expansion. The idea is based on the fact that exchanges are compatible with the required Regge behavior. Hence, to find a complete amplitude, once its singular part is known, we just need to promote each conformal block in the conformal block decomposition of the singular part to the associated exchange, see e.g. \cite{Turiaci:2018dht}. As was reviewed in section \ref{sec:tree}, the coefficient function of the conformal partial wave expansion for the exchange diagram in the direct channel has the form of a product of the standard kinematical $B$ factor times another pole factor
\begin{equation}
\label{2aug1}
A=\frac{\beta}{\nu^2 + (\Delta-h)^2},
\end{equation}
 responsible for generation of singular contributions (\ref{21jun7}). This means that to find the complete amplitude from its singular part given by its $C$-coefficient function, we just need to consider all singularities of $A=C/B$ and for each of them add a term of the form (\ref{2aug1}) with the appropriate locations of poles and the appropriate residues to $A$ of the complete amplitude. Putting differently, Regge behavior bounds translate into bounds on $A$ at $\nu\to\infty$, which allows to reconstruct $A$ from its singularities. 
 
 The resulting sum over poles may require regularization. In practice this boils down to a subtraction of few polynomial terms in $\nu$ with infinite coefficients, which renders the amplitude finite. Related discussions in other representations can be found in \cite{Fitzpatrick:2011dm,Aharony:2016dwx}. It is worth noting that using this approach one can compute individual diagrams and the result does not have to be crossing symmetric.

This idea can be combined with the previously explained approach of computing amplitude's singularities to evaluate any loop amplitude. The main technical difficulty for implementing this approach in practice is that the conformal partial wave expansions for exchanges in the crossed channel are rather complicated. They cannot be computed using the methods reviewed in section \ref{sec:tree}. Instead, one can first compute the conformal partial wave expansion for the exchange in the direct channel and then convert it to the crossed channel employing the crossing kernels or use alternative methods. For recent discussions of conformal block decompositions of exchanges in the crossed channel,  see \cite{Alday:2017gde,Liu:2018jhs,Zhou:2018sfz}. It would be interesting to test the utility of this approach to computing loop amplitudes in practice.

\subsection{Higher-spin theories}

In this section we will consider separately a rather special case of free vector\footnote{Accordingly one should replace $1/N^2$ with $1/N$ compared to the rest of the paper.} models and their higher-spin bulk duals. The approach to the computation of loop corrections in higher-spin theories based on the analytic structure of bulk diagrams was used in 
\cite{Ponomarev:2019ltz} and we would like to provide justifications for some assumptions made there. We refer the reader to \cite{Ponomarev:2019ltz} for the relevant background material on higher-spin theories and vector models. 

From the boundary theory perspective the problem of  $1/N$ corrections may seem trivial. Indeed, given that the theory is free, all correlators can be readily computed and they do not receive any $1/N$ corrections. Naively, one may think that this implies that all bulk loop corrections should  vanish. However, this should not necessarily be the case. The reason is that the identification $G_N = {1}/{N}$ between the bulk Newton's constant and $N$ may be valid only at the leading order in $1/N$. This means that bulk amplitudes may receive loop corrections, but these should be proportional to the tree-level result. Indeed,  if this is the case, the agreement with the boundary result may still be achieved by the appropriate shift in the identification between bulk and boundary coupling constants. By studying vacuum diagrams it was found \cite{Giombi:2013fka,Giombi:2014iua,Giombi:2014yra,Skvortsov:2017ldz} that such a shift is, indeed, necessary. Accordingly, loop corrections for non-vacuum diagrams should also be non-vanishing.  Aiming to confirm this, in \cite{Ponomarev:2019ltz} the double-cut singularity of the one-loop four-point amplitude in the higher-spin theory was computed. Below we will comment on some peculiar issues related to the application of the analysis of previous sections to this case. 

As explained above, we are free to take tree-level four-point functions of a higher-spin theory to be equal to the connected part of the four-point correlator of the $O(N)$ (one can similarly consider $U(N)$ and $USp(N)$ cases) vector model, however, keeping in mind, that the identification between the coupling constants in the bulk and on the boundary may eventually be different from $G_N = {1}/{N}$. To start, we will focus on the scalar four-point amplitude. 
As we are dealing with  four identical fields, the $B$ factor contributes second order poles at double-trace locations to the coefficient function of the conformal partial wave expansion of the four-point tree-level amplitude. In the higher-spin case, given that the boundary theory is free from anomalous dimensions, these singularities should be compensated by  zeros from the $A$ factor, so that the product $A\times B$ has only simple poles. Let us see what this peculiarity of tree-level diagrams implies at loop level.

To construct one-loop diagrams from tree-level ones we proceed as in section \ref{sec:loopfromtree}. This requires to extend some of the external lines of tree-level amplitudes off-shell. We will assume that this extension results in the same analytic structure in $\nu_1$ and $\nu_2$ --- see (\ref{27jun1}) --- as for more standard theories in AdS\footnote{This may be a tricky step considering that holographically reconstructed higher-spin theories are non-local in a conventional sense \cite{Sleight:2017pcz,Ponomarev:2017qab}, hence, in principle, infinite-derivative terms can generate additional singularities. It  would be interesting to see whether such singularities can affect the analysis of section \ref{sec:sing}.}. Carrying out the remaining steps as before, we find that the one-loop amplitude has simple poles at double-trace locations and the associated conformal block coefficients are related to those at tree level by (\ref{28jun4}). In other words, despite a rather peculiar structure of higher-spin theories, (\ref{28jun4}) can still be used to compute the double-trace contributions to the conformal block decomposition of the one-loop diagram with a double-cut in a given channel. Moreover, considering that anomalous dimensions are absent, this gives a complete double-trace part for this amplitude. It is worth stressing, however, that, in contrast to ordinary theories in AdS, these contributions are not singular. Moreover, they  are not specific to diagrams with a non-trivial double-particle cut in a given channel: in particular, these contributions are present already for tree-level four-point functions. This issue complicates the application of the standard unitarity method for the computation of higher-spin amplitudes at one loop. 

A complete computation in the higher-spin theory also requires to take into account contributions from higher-spin fields running in the loop. It turns out that the tensorial structures appearing in tree-level amplitudes are the same as for mean field theory correlators, so one can still use the appropriate generalization of (\ref{28jun4}) as discussed in section \ref{sec:sf}. To summarise, with some reasonable assumptions, the methods presented above can be used to compute a complete double-trace part of all one-loop diagrams with a non-trivial double-particle cut in higher-spin theory. Reconstruction of the complete amplitude is, however, more tricky.

Before concluding, we briefly consider the boundary interpretation of this computation. 
In the higher-spin case the analysis of section \ref{sec:deg} applies except that the anomalous dimensions are vanishing. For scalars, this means that the $O(1/N)$ CFT data induces a $O(1/N^2)$ contribution
\begin{equation}
\label{3jul1x1}
\begin{split}
\langle {\cal O}{\cal O}{\cal O}{\cal O} \rangle^{(2)} \supset 
\left(c^{(1)}_{{\cal O}{\cal O}[{\cal O}{\cal O}]_{n,l}} \right)^2
G^{\Delta}_{\Delta_{n,l}},
\end{split}
\end{equation}
 to the four-point correlator,
which is consistent with the bulk analysis. The connection with the bulk computation gets more tricky if we take into account contributions from all spins. The reason is that the double-trace operators associated with pairs of fields running in the loop mix up in a non-trivial way --- the associated two-point functions are not diagonal. As a result, the bulk summation over all pairs of spins that run in the loop does not seem to have a straightforward counterpart on the CFT side.

\section{Conclusions and outlook}
\label{sec:conc}
In this paper we considered a general one-loop four-point amplitude for scalar fields in AdS admitting a non-trivial double-particle cut in a given channel. By employing the split representation for bulk-to-boundary propagators it was expressed in terms of off-shell tree-level four-point amplitudes. Then, by studying the analytic structure of the resulting spectral integrals we expressed the double-particle singularity of the loop amplitude in terms of the tree-level data. The main result of the paper is given by (\ref{7jun7}) and may be regarded as the AdS counterpart of the flat space formula that relates the discontinuity of a one-loop amplitude associated with a pair of particles going on-shell to tree-level diagrams by unitarity. The analogy is the most transparent if flat space amplitudes are expressed in terms of partial waves with definite spin and energy in the center of mass frame.

Throughout the paper we employed the conformal partial wave expansion for bulk amplitudes. This representation is particularly convenient for establishing the connection with the CFT data  on the boundary. We demonstrated that relation (\ref{7jun7}) translates into a simple statement that the $O(1/N^2)$ CFT data defines a certain singular part of the four-point correlator at order $O(1/N^4)$. This relation was used recently rather extensively both for computing loop diagrams in AdS and $O(1/N^4)$ corrections to correlators in conformal field theories. In this regard, our result shows that this relation can be justified purely from the bulk analysis, that is without resorting to the CFT dual description.

Our findings admit a number of straightforward generalizations that we briefly discuss in section \ref{sec:gen}. In particular, they seem to admit a rather straightforward generalization to higher-point amplitudes and to cuts involving more than two propagators, still giving the results, consistent with the expectations from the large-$N$ considerations on the boundary. Supplemented with the techniques of reconstructing the amplitude from its singular part, these results may be instructive in showing that holography works at any loop order once the duality is true at tree level. 

Finally, let us note that despite the analysis carried out in this paper was perturbative, there are reasons to expect that it can be extended to the non-perturbative level in some way. Indeed, flat space unitarity constrains singularities of the complete non-perturbative $S$-matrix and a similar relation should also be true in AdS. A precise understanding of how this might work is complicated by the difficulties with the definition of multi-trace operators at finite $N$, see \cite{Fitzpatrick:2012yx,Komargodski:2012ek}. It would be interesting to clarify this in future.

\acknowledgments

We are grateful to S. Caron-Huot, P. Kravchuk, E. Sezgin, E. Skvortsov and E.Y. Yuan for valuable discussions on various aspects of the project. This work was supported by NSF grants PHY-1521099, PHY-1803875 and the Mitchell Institute for Fundamental Physics and Astronomy. We also thank Perimeter Institute for hospitality during the conference ''Bootstrap 2019'' where this work was completed. Research at Perimeter Institute is supported in part by the Government of Canada through the Department of Innovation, Science and Economic Development Canada and by the Province of Ontario through the Ministry of Economic Development, Job Creation and Trade.

\appendix
\section{Singularity structure: an example}
\label{App:Ex}

In this Appendix we consider a toy example of an integral that shares many common features with
 the integral we studied in section  \ref{sec:sing} and, at the same time, can be easily evaluated exactly. We give this example to illustrate that the contour deformation that we used in the main text does allow to capture the singularities of the original integral correctly, at the same time, producing additional shadow poles.  A systematic account of the topic can be found in \cite{smatrix,Yuan:2018qva}.

Consider an integral
\begin{equation}
\label{11jun1}
I(\nu)=\int_{-\infty}^\infty d\nu_1d\nu_2 f(\nu_1,\nu_2,\nu),
\end{equation}
where
\begin{equation}
\label{11jun2}
\begin{split}
& f(\nu_1,\nu_2,\nu)=\frac{(h^2+\nu^2)\nu_1^2 \nu_2^2}{\big( (\Delta_1-h)^2+\nu_1^2\big) \big((\Delta_2-h)^2+\nu_2^2\big)}\\
&\quad\frac{1}{\left((\nu+\nu_1+\nu_2)^2 + h^2\right) (\left  (\nu-\nu_1+\nu_2)^2 + h^2\right) (\left (\nu+\nu_1-\nu_2)^2 + h^2 \right) (\left (\nu-\nu_1-\nu_2)^2 + h^2 \right)}.
\end{split}
\end{equation}
This integral has the analytic structure similar to that of (\ref{6jun12}), except that we replaced everywhere gamma functions, producing series of poles or zeros, with single poles or zeros at locations where the arguments of the respective gamma functions vanish. 
 Indeed, besides the explicit propagator factors (\ref{27jun3}), in (\ref{11jun2}) we also have poles at locations of the leading singularities generated by the gamma functions in the numerator of (\ref{27jun5}). In addition (\ref{11jun2}) has zeros $\nu^2_1\nu_2^2$ as in (\ref{6jun12}) and $(h^2+\nu^2)$ instead of two series of zeros from $(\Gamma(h+i\nu)\Gamma(h-i\nu))^{-1}$ in the explicit prefactor in (\ref{6jun12}).

The integral (\ref{11jun1}) can be evaluated exactly using the residue theorem two times. The result is
\begin{equation}
\label{11jun3}
I(\nu)=\frac{\pi^2(\Delta_1+\Delta_2)}{16 h \Delta_1\Delta_2\big(\nu^2 + (\Delta_1+\Delta_2-h)^2\big)}.
\end{equation}
We can see that the integral has  poles only at locations that can be regarded as double-trace locations (\ref{27jun6}) with $n=0$,
that is 
\begin{equation}
\label{11jun3x2}
h\pm i\nu = \Delta_1+\Delta_2.
\end{equation}

To capture these singularities of $I$, one can instead consider an integral of $f$ along contours encircling singularities of the integrand generated by the propagator factors. The new integral can be evaluated by collecting the residues at $\nu_1=\pm i(\Delta_1-h)$ and $\nu_2=\pm i(\Delta_2-h)$
 with the result
\begin{equation}
\label{11jun3x1}
\begin{split}
I'(\nu) &\equiv 4(2\pi i)^2 \res_{\nu_2=-i(\Delta_2-h)} \res_{\nu_1=-i(\Delta_1-h)} f(\nu_1,\nu_2,\nu)\\
&
\qquad\qquad=\frac{4\pi^2 (h^2+\nu^2)(\Delta_1-h)(\Delta_2-h)}{\big(\nu^2+(\Delta_1+\Delta_2-h)^2\big)
\big(\nu^2+(\Delta_1+\Delta_2-3h)^2\big)
}\\
&\qquad\qquad\qquad\qquad\qquad\qquad
\frac{1}{\big(\nu^2+(\Delta_1-\Delta_2-h)^2\big)
\big(\nu^2+(\Delta_2-\Delta_1-h)^2\big)}.
\end{split}
\end{equation}
It is straightforward to verify that
\begin{equation}
\label{2jul1}
\res_{\nu=\pm i(\Delta_1+\Delta_2-h)} I(\nu)= \res_{\nu=\pm i(\Delta_1+\Delta_2-h)} I'(\nu),
\end{equation}
as required. In other words, $I'$ correctly captures singularities of the original integral $I$. However, it is not hard to see that $I'$ has additional poles not present in $I$.

This example can be extended in various ways to illustrate the features we encountered in the main text. For instance, if we assume that the integrand itself has poles in $\nu$ at locations (\ref{11jun3x2})  then, clearly, the integral will have poles at these locations of order higher by one.

\bibliography{loops1N}
\bibliographystyle{JHEP}

\end{document}